\documentstyle[epsf]{mn}

\nonstopmode	

\def\etal{{\it et al.\hskip 3pt}}
  
\def\ie{{\it i.e.\hskip 3pt}}
\def\eg{{\it e.g.\hskip 3pt}}
\def\plotone#1{\centering \leavevmode
\epsfxsize=\columnwidth \epsfbox{#1}}
\def\plottwo#1#2{\centering \leavevmode
\epsfxsize=.45\columnwidth \epsfbox{#1} \hfil
\epsfxsize=.45\columnwidth \epsfbox{#2}}

\title[Filament and Shape Statistics]
{Filament and Shape Statistics: A Quantitative Comparison of Cold + Hot and
Cold Dark Matter Cosmologies vs. CfA1 Data}

\author[R. Dav\'e \etal]{Romeel Dav\'e$^1$, Doug Hellinger$^2$,
        Joel Primack$^2$, Richard Nolthenius$^1$, and Anatoly Klypin$^3$ \\
$^1$ UCO/Lick Observatory, University of California, Santa Cruz, CA 95064 \\
$^2$ Santa Cruz Institute for Particle Physics, University of California, Santa Cruz, CA 95064 \\
$^3$ Astronomy Department, New Mexico State University,
     Las Cruces, NM 88001 }

\date{Accepted ???; Received ???}
\pubyear{1996}

\begin{document}

\maketitle

\label{firstpage}

\begin{abstract}
A new class of geometric statistics for analyzing galaxy catalogs is presented.
{\it Filament statistics} quantify filamentarity and planarity in large
scale structure in a manner consistent with catalog visualizations.
These statistics are based on 
sequences of spatial links which follow local high-density
structures.  From these link sequences we compute
the discrete curvature, planarity, and torsion. Filament statistics are applied 
to CDM and CHDM ($\Omega_\nu = 0.3$) simulations of Klypin \etal (1996),
the CfA1-like mock redshift catalogs of Nolthenius, Klypin and 
Primack (1994, 1996), and the CfA1 catalog.
We also apply the moment-based shape statistics developed 
by Babul \& Starkman (1992), Luo \& Vishniac (1995),
and Robinson \& Albrecht (1996) to these
same catalogs, and compare their robustness and discriminatory
power versus filament statistics.
For 100 Mpc periodic simulation boxes ($H_0 = 50$ km s$^{-1}$ Mpc$^{-1}$),
we find discrimination of $\sim 4\sigma$ (where $\sigma$ represents
resampling errors) between CHDM and CDM for selected filament statistics
and shape statistics, including variations in the galaxy identification scheme.
Comparing the CfA1 data versus the models does not yield a conclusively
favored model; no model is excluded at more than a $\sim 2\sigma$ level 
for any statistic, not including cosmic variance which could further 
degrade the discriminatory power.
We find that CfA1 discriminates between models poorly mainly due to its 
sparseness and small number of galaxies, not due to redshift distortion,
magnitude limiting, or geometrical effects.  
We anticipate
that the proliferation of large redshift surveys and simulations
will enable the statistics presented here
to provide robust discrimination between large-scale structure in 
various cosmological models.
\end{abstract}


\section{Introduction}

In this paper
we develop and apply statistics to quantitatively characterize the shapes
of galaxy distributions seen in redshift surveys.  We then use these
statistics to compare cosmological simulations of pure Cold Dark 
Matter (CDM) models versus Cold plus Hot Dark Matter (CHDM) models
in real space, as well as
simulated CfA1-like redshift surveys generated from these
simulations versus the CfA1 data.
The four simulations used here are summarized in Table~\ref{table: sims}
and described in detail in \S \ref{sec: sims}.
The availability of this suite of simulations of different cosmological
models, all computed and analyzed in parallel, allows us to test
the ability of these statistics to discriminate between such models.
Visual comparison of the simulations (Brodbeck \etal 1996; hereafter BHNPK) 
shows that the CDM galaxy distribution contains larger clusters and
less well-defined filamentary and sheet-like structures than CHDM,
consistent with the fact that CHDM forms structure at a later epoch than CDM.
The statistics presented here
confirm as well as quantify these results, showing
statistically significant and robust discrimination between the models.

Ever since the CfA1 Survey (deLapperant, Geller \& Huchra 1988) detected 
filamentary and planar structures in the galaxy distribution,
many attempts have been made to develop statistics computed
solely from the redshift-space positions of galaxies which
quantify these large-scale structures.  It became apparent that 
the two-point correlation function contains very limited information
about structure, while higher-order correlation functions are
difficult to measure and mainly decribe the highest density regions.  
Thus alternative, more geometrical
methods have been developed which contain information from {\it all} orders of correlation
functions in such a way as to characterize the types of structures seen in 
the surveys.  The void probability function ({\it e.g.}
Vogeley, Geller \& Huchra 1991,  Ghigna \etal 1994, 1996)
and the topological genus statistic (see Melott 1990 for review, and
Coles, Davies \& Russell 1996 for related discussion) 
have had some success,
and lately more complex statistics have been developed which seem promising
such as the Minkowski Functionals (Mecke, Buchert \& Wagner 1994 and
Kerscher \etal 1996).

Here we introduce {\it filament statistics}, a new class of geometric 
statistics designed to quantify filamentarity and planarity in 
large-scale structure.  Filament statistics
use information about the moments of the local mass
distribution to characterize the shape of large-scale structure.
In this way they are similar to the shape statistics of 
Babul \& Starkman (1992; hereafter BS), 
Luo \& Vishniac (1995; hereafter LV), and
Robinson \& Albrecht (1996; hereafter RA).
However, the way in which the moment tensor information is 
used in filament statistics is fundamentally different from these
shape statistics.  Rather than randomly sampling the galaxy distribution,
filament statistics use a prescription to map the galaxy distribution
into a new set of points which amplifies the properties showing the
greatest differences between models, namely filamentarity and planarity.
While statistics applied to the new point set can be more discriminatory, 
care must be taken to develop a prescription which is robust
against inherent variations in the simulated galaxy distribution such as
galaxy identification uncertainty and cosmic variance.
Hence in constructing filament statistics,
we are guided by the following principles:

\begin{itemize}

\item 
It is best to attempt to directly quantify structures which visually show
the greatest differences between models, {\it i.e.} filaments and sheets.

\item 
It is best to apply statistics directly to the point set of galaxies
rather than a smoothed density distribution to avoid discarding
information on small scales.

\item 
It is best to construct simple, interpretable statistics, 
in order to more easily understand their robustness against 
intrinsic uncertainties in the analysis.

\end{itemize}

We find that both our filament statistics 
and the BS, LV and RA shape statistics yield statistically significant
discrimination between CDM and CHDM simulations, which persists 
(though to a significantly lesser degree) even in a redshift-space comparison 
versus the CfA1 data.
A more informative comparison of these statistics must await the
availability of larger, more complete redshift surveys
as well as simulations capable 
of properly modeling these large volumes of space; both should
be available soon.
For now, we present these results to demonstrate the viability of the methods.  
We note that this is the
first publication in which any of these moment-based shape statistics 
have been used to compare simulations to
redshift survey data, which is the purpose for which they
were originally devised.

\section{Implementation of Filament Statistics}

\subsection{Dekel's Alignment Statistic}

Our filament statistics are related to the alignment statistic, originally proposed
by Dekel (1984):  For each galaxy, consider
two concentric shells; find the moment of inertia ellipsoid axes defined by galaxies
within each shell; and calculate the angle difference between the
inertia tensor axes.  Presumably, where the
angle difference in the major axis is small, there is a filamentary
structure present, and where the angle difference in the minor axis
is small, there is a sheet-like structure present.  By randomly sampling
the galaxy distribution at different shell radii, one can then 
gain a measure of the filamentarity
and planarity in large-scale structure at various scale lengths.  
However, we found that the alignment statistic barely discriminated between CDM and CHDM
in real 3D space, and failed to discriminate the models in redshift space.

Since the visualizations of BHNPK show marked differences in
the number, size, and continuity of filamentary structures in CHDM and CDM,
we were inspired to consider mapping the 
point set of galaxies into another point set 
by an algorithm that sensitively favors contiguous high density regions.

\subsection{The Creation of Link Sequences}

The basis of filament statistics 
is the creation of {\it link sequences} which follow along
local high density regions, as determined by the principal axis
of the local moment of inertia tensor.  
Link sequences may be thought of as a technique to map
the point set of galaxies into a new point set which emphasizes the 
higher density regions containing the filamentary and planar structures
which we would like to quantify.  A statistic applied to this new
point set is likely to be more discriminatory than the same statistic 
applied to a random subset of galaxies; this is the case with
the alignment statistic.
Most other statistics presented in the literature (including BS, LV, and
RA statistics) use sampling of the data points to obtain a measure of the global
structure; filament statistics represents a new method for 
manipulating the data to enhance the structures of interest, 
thereby enhancing the discriminatory power of a given statistic.

A link sequence is an ordered set of points which can be visualized as
joined by ``links", created by the procedure outlined in the flowchart
in Figure~\ref{fig: dg_flowchart_filament}.
A link sequence is started from each galaxy in a catalog of galaxies
(or if there are too many, a random subset of such galaxies).
The moment of inertia tensor is computed using the masses and positions 
of galaxies within a range $R$ of the given point; for redshift survey data, 
we weight by luminosity instead of mass.
The eigenvectors and eigenvalues of the inertia tensor are found, and from these
the principal axis is determined.  
The new point in the sequence is created at a distance $L$ (the ``link length")
away in the direction of the principal axis, and a link is created
which joins the old point to the new point.
Note that only the first point in a link sequence is a galaxy; the others
are simply locations within the catalog volume. 
A new inertia tensor is computed around
this new point, and the procedure is repeated until termination.  
Sequence termination occurs when there
are too few nearby galaxies to reasonably identify an axis.  
By this prescription, each galaxy generates a sequence of links. 
If a sequence has more than  $N_{L,min}$ links,
then statistics are computed on this link sequence, otherwise 
the sequence is discarded.
The construction of a link sequence is completely defined by choosing the 
link length $L$,
the maximum radius of galaxies to be included in computation of the 
moment tensor $R$, and the criteria for termination of a sequence.

Note that the principal axis of the inertia tensor does not define a
unique direction.
So from the initial point, the sequence is propagated in both (opposing)
directions until termination, and the entire joined sequence is
what is used for statistical analysis, as long as the total number
of links is at least $N_{L,min}$.
Generally, sequences tended to be non-intersecting but in some cases they 
oscillated between two points.  When this is detected, the
sequence is terminated.  


\subsection{Constructing a Dimensionless Statistic}

We would like to construct dimensionless parameters which describe the
shapes of structures.  For that we need to express all scales in
units of some typical length scale of the catalog.
A natural choice is the mean intergalaxy spacing 
$\bar d \equiv (V/N)^{1/3}$, where $V$ is catalog volume and $N$
is number of galaxies in the catalog, since it provides
a length scale independent of galaxy clustering; it is
also the simplest choice.  

We will consider applications in real space as well as magnitude-limited
and volume-limited redshift space.
Redshift distortion will produce measurable effects on link sequences,
and attempts will be made to understand and quantify these effects.
Whereas in real space the mapping of a galaxy into a link sequence
is completely well-defined, in redshift space this is no longer true ---
redshift distortion for a given structure depends on the
vantage point chosen to observe the structure.
In this paper we introduce a new method for quantifying how
redshift-space distortion affects these statistics.  
We show that none of the statistics analyzed here are adversely affected by
redshift distortion to any significant degree.

A complication arises in computing $\bar d$ in magnitude-limited catalogs,
since the sample incompleteness increases with distance from the Milky Way
origin, making $\bar d$ a function of radius from origin.  
A local computation of $\bar d$ around a given sequence point 
({\it i.e.,} using $\bar d = (V/N)^{1/3}$ for a local volume $V$
around the given point) will degrade the statistics, 
since structure identification will be biased
towards underdense regions where $\bar d$ is large, which is
exactly opposite of what is desired.  Instead, $\bar d$ should be 
corrected using the selection function,
which depends only on the distance
$r$ from the origin.  Since $\phi (L) dL$ is the number density of galaxies
between luminosity $L$ and $L+dL$, we can obtain $\bar d (r)$
for galaxies visible above the magnitude limit as follows:

\begin{equation}
\bar d (r) = \left\lbrack {\int^\infty_{L_{\em lim}(r)} \phi (L) dL} \right\rbrack^{-1/3}
\label{eq: dgmis}
\end{equation}

\noindent
where $L_{\em lim}(r)$ is the luminosity
of a galaxy with apparent magnitude 14.5 (the CfA1 magnitude limit) at
a distance $r$.  $\phi (L)$ is assumed to have Schecter form

\begin{equation}
\phi (L) dL = \phi^\ast \biggl({L\over L^\ast}\biggr)^\alpha \exp(-L/ L^\ast) {dL\over L^\ast}
\end{equation}

\noindent
The Schecter parameters $\phi^\ast$, $L^\ast$ and $\alpha$ 
were best-fit to each real and simulated redshift catalog 
individually; this procedure is described in Nolthenius \etal 
(1994, 1996; NKP94 and NKP96, respectively).  
Note that the true distance
$r$ is unknown, and is instead estimated assuming no peculiar
velocities, {\it i.e.} $r \equiv v/H_0$ for a galaxy with radial velocity $v$.  
In the CfA1 data, a few blueshifted galaxies 
(mostly in Virgo) do end up on the opposite side of the origin, but the
statistics turn out to be insensitive to where these few galaxies are placed.  
${\bar d (r)}$ is computed and used as the local
mean intergalactic spacing at each sequence point in the
analysis of magnitude-limited catalogs.

\subsection{Link Parameters}

The first parameter choice we tried was the simplest, with
$L =R = \bar d$.  The virtue of this definition
is that we have a parameter-free statistic, in the sense that the
parameters are all determined from intrinsic properties of the data set.
Unfortunately, statistics derived from constructions with these 
``natural'' parameters did not discriminate between models.
$R = \bar d$ turns out to be 
too small to identify a local structure, and is dominated by shot noise.

For link length $L = \bar d$, but range $R$ left as a free parameter, we obtain
discriminatory statistics; this choice of $L$ appears to work as well as
any other.  However, for $R = \bar d$, and $L$ a free parameter, 
we again find little discrimination between models, or
even from a Poisson catalog.  
A larger $R$ will yield more points per sphere, thereby lowering shot-noise scatter.
Since the $R$ parameter controls the scales of structure being
measured by the statistics, it is interesting and instructive to look at statistics as
a function of $R$, and the results will be presented that way.


\subsection{Termination Criteria}

There are three parameters which set the termination criteria for a 
link sequence.  $N_{P,min}$ is the minimum number of galaxies required within a
sphere of radius $R$ for a sequence to continue; $N_{P,min}$ was set to
5 so that the determination of the principal axis would be statistically
meaningful, and so that a sequence would terminate if it was in a
sparse region in the catalog.
$N_{L,max}$ sets the
maximum number of links for a periodic catalog,  and
is set so that the total length of a sequence cannot exceed the length
of the simulation box.  In a redshift survey, the sequence terminates if
it exceeds the catalog boundary.
$N_{L,min}$ sets the minimum number of links for
a sequence to be statistically meaningful.  This was set to 4 links 
(the minimum value for
computation of all statistics), but can be increased to explore
more extended structures.  However, since each link is typically
fairly large ($\approx$3 Mpc in the simulations considered, and 
$\approx$15 Mpc in the sparser CfA1 catalog), 4 links is already 
exploring a reasonably extended scale.

All the termination parameters were varied over fairly wide ranges. 
$N_{P,min}$ was varied from 4 to 10 with little change in discrimination
or robustness; any higher, and the shot noise generated from fewer
sequences became significant.  The statistics are
independent of $N_{L,max}$ as long it is above about 10, below which
shot noise from the small number of links becomes significant; it is
$\sim 34$ in the periodic simulation boxes.  Variations in $N_{L,min}$ 
had some effect on the results for real and simulated redshift catalogs,
since for very small values (2 or 3)
shot noise increases from low link sampling, while for a high value (above
10), the number of acceptable sequences decreases so that catalog sampling
shot noise becomes large.
Discrimination was also insensitive to the choice of either a Gaussian, exponential,
or top hat window function;  we used a top hat for computational efficiency.


\subsection{Computation of Statistics}

Once link sequences have been generated, there are two ways the
new point set may be used.  We may choose to apply statistics
which were previously applied to random subsets of the data to this 
new point set.  Alternatively, we may devise statistics which
measure the properties of link sequences themselves.  Filament
statistics are based on the latter idea, following the intuitive
characterizations of structure given by the alignment statistic.

We developed three statistics to compute on a link sequence
which measure filamentarity or planarity in an easily interpretable way.
We call them planarity, curvature, and torsion.
These statistics are in general defined as angle deviations between
inertia ellipsoid axes for consecutive points along a link sequence; the exact
definitions are as follows:

\begin{itemize}

\item
{\it Planarity} ($\theta_P$)
is the angle difference between the minor axis of the inertia tensor for two
consecutive points.
The geometrical interpretation of planarity is as follows:
Given that filaments in large-scale structure often
occur at intersections of sheet-like structures, the minor axis of
the inertia tensor along the filament measures the strength of the 
embedding sheet perpendicular to the filament; 
hence a lower planarity angle indicates the presence of a 
local sheet-like structure.

\item
{\it Curvature} ($\theta_C$) is defined as the angle difference between two
consecutive links.  Equivalently, it is the angle difference between the major axis
of the inertia tensor for two consecutive points.  
A sequence which is following a well-defined filament will have a low
angle difference between links;  hence a lower curvature angle indicates
greater filamentarity.  

\item
{\it Torsion} ($\theta_T$) is
the angle difference between the plane defined by the 
first two links and the third link.  Torsion measures the
strength of the embedding sheet parallel to the filament, a lower
torsion indicating a stronger planar structure present.

\end{itemize}

In all cases, {\it a lower value (angle difference) signifies more 
structure present in the catalog}.  As an example, consider a
set of points distributed randomly throughout a long, thin cylinder.
A sequence will track the cylinder, and the angle deviation between
each successive link will be very small; hence curvature will show
a very low angle deviation.  Conversely, planarity and torsion will show 
large angle deviations since there is no locally preferred plane
in a circular cylinder.
For a thin sheet, sequences will randomly walk throughout the sheet,
yielding a high curvature angle (indicating no filamentary structure), but
low planarity and torsion angles (indicating lots of planar structure).

In large-scale structure, filaments are often embedded within
sheets, and thus these statistics are expected to be
correlated.  Nevertheless it is useful
to consider each one separately.  A key difference between
the statistics is that each requires a different number of sequence points
to compute.  Planarity is the most {\it local} statistic, being computed
from only 2 link nodes, while curvature requires 3, and torsion
requires 4.  While planarity and torsion are in the ideal case purely
measures of planarity, torsion is more sensitive to the presence of
local filamentary structure since it measures angle
differences along the sequence rather than perpendicular to the sequence.

For each of those statistics, an average value
is found within a single sequence.  Then, for all the
sequences in that catalog, a median
value is found.  We will denote the resulting averaged-then-medianed 
statistic by a bar,
as in $\bar\theta_C$.  This final median value is the value of that statistic
for the given catalog at the selected value of $R$.  
Errors analysis is discussed in section 4.

\subsection{Visualization and Algorithm Testing}

We have attempted to construct an algorithm which will identify and
track filaments.  We tested the algorithm on artificially generated
point sets of lines and planes of varying thickness.  The results
conformed to qualitative expectations, that lines should show a
great deal of filamentarity and little planarity, and vice
versa for planes.  Also, the median angle deviations increased
with thickness, as expected.  Visualizations showed that link sequences
were tracking the structure as expected.

When we visualized the link sequences which were generated in an
actual CHDM simulation,  they
tended to lie preferentially in regions of structure,
but could not often be associated with visually
recognizable filaments.  They were also scattered throughout
the simulation volume.
This is because for the simulations we considered (which will be
described in the next section), nearly every galaxy that was tried
as a sequence starting point yielded a qualifying ($N_{L,min}\geq 4$)
sequence.  Thus the parameter set we have chosen does not sufficiently
restrict the generated sequences to lie directly along the
filaments that are detected by eye.  By imposing more severe
requirements for sequence qualification, one can tune the
algorithm to better recognize filamentary patterns.  However, this reduces
the number of qualifying sequences to a point where statistics
are poor, and hence it is not useful for performing 
statistically significant comparisons.
Our conclusion is that this algorithm is not particularly 
suited for pattern recognition, and
is better suited for statistical comparison of overall structural
properties of models.
The statistics we compute have simple interpretations, and the
results for various models are
consistent with the BHNPK visualizations;  however, this agreement is not
necessarily apparent from visualizations of individual link sequences.

Little effort went into developing analytical predictions for
expected values of $\bar\theta_C$, $\bar\theta_P$, and $\bar\theta_T$,
even in the case of a Poisson catalog.
This is due primarily to the fact that the algorithm was
successful in the test cases we considered, and thus a 
complex and time-consuming analytical prediction was deemed
to be low priority.  Further numerical testing may also be done
by superimposing lines or sheets of varying strengths on a Poisson catalog,
and determining how effective the algorithm identifies structure.
We leave these endeavors to the future, and instead for now
concentrate on applications to the comparison of cosmological models.

\section{Moment-Based Shape Statistics}

As a brief review, we present the definitions of statistics given
by BS, LV, and RA.  Since one of the authors independently
derived the LV statistics (Hellinger 1995), we present those first
and in somewhat more detail.
The construction of each family of statistics is well-motivated and 
elegantly presented in the relevant papers; we refer the reader to them
for further details, and here focus on the definitions.

\subsection{Luo \& Vishniac Shape Statistics}

The statistics presented in LV are a three-dimensional extension of the
two-dimensional shape statistics devised by Vishniac (1986).
In three dimensions, LV statistics are given by a linear combination
of the quadratic coordinate moment invariants
(summations implied by repeated indices):
\begin{eqnarray}
     LV & = &  C_1 ({\cal M}^{ii})^2 
        + C_2 ({\cal M}^{ij})^2 
	+ C_3 ({\cal M}^{ii}) ({\cal M}^j)^2 \nonumber \\ & &
        - C_4 ({\cal M}^{ij}) ({\cal M}^i) ({\cal M}^j),
\end{eqnarray}
where
\begin{eqnarray}\label{eqn: moments}
{\cal M}^i & \equiv & {1\over M}\sum_k m_k (x_k^i - x_0^i)  \nonumber \\
{\cal M}^{ij} & \equiv & {1\over M}\sum_k m_k (x_k^i - x_0^i)(x_k^j - x_0^j)  \nonumber \\
M & \equiv & \sum_k m_k
\end{eqnarray}
are summed over all galaxies at ${\bf x}_k$ with masses $m_k$
within a window radius $R$ of a central galaxy at ${\bf x}_0$.
The constants $C_i$ are determined by the constraints applicable for a given
shape.  For a ``filamentarity'' statistic we have
$$
\left\{
\begin{array}{ll}
	LV \equiv 0 \mbox{ for a spherical distribution entirely within} \; R \\
	LV \equiv 0 \mbox{ for a distribution with a uniform gradient,} \\
	\; \; \; \; \mbox{of arbitrary size, in the window defined by} \; R \\
	LV = 1 \mbox{ for a uniform linear density passing } \\
	\; \; \; \; \mbox{through the window center}
\end{array}
\right.
$$
This yields the {\it quadratic filamentarity} statistic:
\begin{eqnarray}
	LV_{\sl quad} &  = & \frac{1}{({{\cal M}}^{ii})^2} 
		\{
		- \frac{1}{2} ({\cal M}^{ii})^2 + \frac{3}{2} ({\cal M}^{ij})^2 \nonumber \\ & &
		+ \frac{1}{2} ({\cal M}^{ii}) ({\cal M}^j)^2 
		- \frac{3}{2}({\cal M}^{ij}) ({\cal M}^i) ({\cal M}^j) \}
\end{eqnarray}
appropriate for comparison of three dimensional real and redshift data.

The quadratic statistic has the virtue of being a lowest order nontrivial
moment invariant shape statistic, and thus can generally be expected to
yield the strongest signals, but it correspondingly has the weakest ability
to discriminate among different clustering shapes.  For example, consider a data set
where all of the galaxies are coplanar.  Here we find that $LV_{\sl quad}=1/4$,
indicating that a purely planar structure can imprint a weak signal; thus the
quadratic statistic cannot fully distinguish lines from planes.  Either
we need to supplement this statistic with a complementary diagnostic,
or we must sacrifice some signal strength and go to a higher order statistic.
We follow LV and choose the latter method.
This gives the following LV cubic structure statistics:
\begin{eqnarray}
	{LV_{\sl line}} & = & \frac{1}{({\cal M}^{ii})^3} 
	\{
	-{\cal M}^{ii}({\cal M}^{ij}-{\cal M}^i{\cal M}^j)^2 \nonumber \\ & &
 +\frac{1}{2}{\cal M}^{ii}({\cal M}^{kk}-({\cal M}^j)^2) \nonumber \\ & &
 +3{\cal M}^{ij}({\cal M}^{jk}-{\cal M}^j{\cal M}^k)
		({\cal M}^{ik}-{\cal M}^i{\cal M}^k) \nonumber \\ & &
 	-\frac{3}{2}{\cal M}^{ij}({\cal M}^{ii}-{\cal M}^{k}{\cal M}^k)
		({\cal M}^{ij}-{\cal M}^i{\cal M}^j) 
	\} \label{eq:Linearity} \\ \nonumber \\
	{LV_{\sl plane}} & = & \frac{1}{({{\cal M}}^{ii})^3} 
	\{
	+4{\cal M}^{ii}({\cal M}^{ij}-{\cal M}^i{\cal M}^j)^2 \nonumber \\ & &
 -4{\cal M}^{ii}({\cal M}^{kk}-({\cal M}^j)^2) \nonumber \\ & &
 -12{\cal M}^{ij}({\cal M}^{jk}-{\cal M}^j{\cal M}^k)
		({\cal M}^{ik}-{\cal M}^i{\cal M}^k) \nonumber \\ & &
 	+12{\cal M}^{ij}({\cal M}^{ii}-{\cal M}^k{\cal M}^k)
		({\cal M}^{ij}-{\cal M}^i{\cal M}^j) 
	\} \label{eq:Planarity}
\end{eqnarray}
properly discriminating linear from planar structures.
We also considered ``flatness'', an equally weighted combination of
linearity and planarity:
$LV_{\sl flat} = {1\over 2}(LV_{\sl line} + LV_{\sl plane})$.  We thought
the flatness statistic may be useful given that CHDM models tend to
show both higher planarity and filamentarity than CDM models, but
in final analysis it was quite similar to $LV_{\sl plane}$, so
we don't consider it separately here.

\subsection{Babul \& Starkman Shape Statistics}

The shape statistics presented in BS
are derived from functions of the moment-of-inertia tensor $I^{ij} \equiv M^{ij}-M^i M^j$,
where $M^{ij}$ and $M^i$ are defined in equation~\ref{eqn: moments} as averages of
coordinate moments within a window of a specified radius $R$.
Following the scheme introduced by Vishniac (1986),
they define three {\it structure functions}:
\begin{eqnarray}
BS_{\sl prol} & = & \sin( \frac{\pi}{2}(1-\nu)^p ) \; \; \; \mbox{``PROLATENESS''} \\
BS_{\sl obl} & = & \sin( \frac{\pi}{2}a(\mu,\nu) ) \; \; \; \; \; \mbox{``OBLATENESS''} \\ 
BS_{\sl sph} & = & \sin( \frac{\pi}{2}(\mu) ) \; \; \; \; \; \; \; \; \; \; \mbox{``SPHERICITY''}  
\end{eqnarray}
where $\mu = \sqrt{I_3/I_1}$, $\nu = \sqrt{I_2/I_1}$,
${I_1 \geq I_2 \geq I_3}$ are the eigenvalues of $I^{ij}$,
$p=\frac{\log(3)}{\log(1.5)} \approx 2.71$, and $a(\mu,\nu)$ is defined implicitly by
$$
\frac{\nu^2}{a^2} - \frac{\mu^2}{a^2(1-\alpha a^{\frac{1}{3}} 
+ \beta a^{\frac{2}{3}})} \equiv 1
$$
with $\alpha=\frac{13 (1+3^{\frac{1}{3}})-{3^{\frac{2}{3}}}}{16} \approx 1.854$,
and $\beta=-\frac{7}{8} 9^{\frac{1}{3}}
+ \alpha 3^{\frac{1}{3}} \approx 0.854$.
The form of the structure functions and the values were chosen to
give functions which are flat near the value of unity for a given morphology
then fall to zero more sharply, reaching 0.5 at an axis ratio of 1:3.
In our case, prolateness quantifies filamentarity, oblateness quantifies
planarity, and sphericity quantifies the clumpiness of the galaxy distributions.

BS found that these statistics could discriminate between
cosmological simulations having Gaussian
random initial perturbations with varying power-law indices.
These statistics were also recently combined with a percolation analysis
and applied to various toy models of structure formation (Sathyaprakash, 
Sahni \& Shandarin 1996).

\subsection{Robinson \& Albrecht's Statistic}

In a recent paper (RA) a combination
of inertia tensor eigenvalues was devised which yields a value of 1 for planar
structures and 0 for filamentary structures.  In RA it is called ``flatness",
but by the LV nomenclature it is actually a planarity measure, since
it gives 0 for filaments.  It is given by:
\begin{equation}
RA = {{\sqrt{3}(I_2 - I_3)\sqrt{I_1^2+I_2^2+I_3^2}\over
I_1^2+I_2^2+I_3^2+I_1 I_2+I_1 I_3+I_2 I_3}}
\label{eq: ra}
\end{equation}

RA found that this statistic was able to distinguish sheet-like non-Gaussianity
in various toy models of cosmic string wakes.

\subsection{Test Cases and Structure Aliasing}

To test our implementation and better understand the quantitative
behavior of these statistics, we apply them to the following three
test cases, each one within a 100 Mpc periodic box:

\begin{itemize}
\item  ``Line" -- 1000 points randomly placed along a single line
extending across the entire volume.

\item ``Plane" -- 5000 points randomly placed in a single plane
extending across the entire volume.

\item ``Sphere" -- 5000 points randomly placed in a spherical distribution
of 5 Mpc radius at the center of the box.
\end{itemize}

We compute the statistics on each of these test cases, with
window radius $R$ = 5 and 10 Mpc.  We show the results,
along with the analytical value for a
continuous distribution (or equivalently, the predicted value
for $R\rightarrow\infty$), in Table~\ref{table: test}.

In general, all computed values agree quite closely with the analytical value
for all statistics.  As $R$ increases, the value approaches the
predicted value, as expected.  Note that the quadratic filamentary 
statistic $LV_{\sl quad} \approx 0.25$ for a plane, indicating (as mentioned
before) that this statistic does not completely distinguish lines
from planes.  Also, note that $RA$ approaches unity at a slower rate
than other planarity measures, and concurrently yields a weak signal for
a discrete sphere.

The deviation from the analytical value is due to discreteness effects
in these test case catalogs, an effect which becomes even more significant
in the sparser sky catalogs.  As an example, consider a window radius
so small it only encompasses 3 points out of the Sphere catalog.  This
configuration will yield a strong planar signal despite the topology
of the underlying distribution.  Analogously, 3 points within a planar
structure can yield a strong linear signal if those three points 
happened to be somewhat colinear.  We call this effect 
{\it structure aliasing}, and it primarily important in lower density
regions, and for smaller window radii.  In the simulations and
redshift surveys, we would like to probe small-scale structure where
the differences between models are greatest, but we are hampered
by increased shot noise and structure aliasing.  Hence we vary
$R$ to determine the optimal scale for discrimination, as
well as to explore the behavior of the statistics at different scales.

\section{The Simulations and Data}

\subsection{The Halo Catalogs}\label{sec: sims}

All statistics were applied to the simulations
described in Klypin, Nolthenius \& Primack (1996; KNP96),
which are 100 Mpc$^3$ particle-mesh simulations
on a 512$^3$ force resolution grid.  All had $\Omega=1$ and $H_0 = 50$ 
km s$^{-1}$ Mpc$^{-1}$ (which will be assumed throughout).
A resolution element, or cell, is 195 kpc.
The CDM simulations had 256$^3$ particles, while the CHDM simulations
had 256$^3$ cold particles and $2 \times 256^3$ hot particles, giving
a cold particle mass of $2.9\times 10^9 M_{\odot}$ and $4.1\times 10^9 M_{\odot}$
for CHDM and CDM, respectively.  There were two simulations with pure
CDM, one with linear bias factor $b=1.0$ (CDM1) and one with $b=1.5$ (CDM1.5), and two
CHDM simulations with $10\%$ baryons, $30\%$ in a single neutrino species and
the rest cold dark matter. 

Both CHDM simulations have linear bias factors which are
compatible with the COBE DMR results, while CDM1 is nearly 
compatible, requiring some tensor contribution.
CHDM$_1$ and both CDM simulations were started with identical random number
sets describing the initial perturbation amplitudes.
It was found in NKP94 and
KNP96 that Set 1 had, by chance, an unusually high power
($\sim \times 2$) on scales comparable to the box size.  However,
the CfA1 data appears to show similarly unusual power when
compared to the larger APM survey data (NKP96, Vogeley \etal 1992, Baugh \&
Efstathiou 1993). CHDM$_2$ had a power spectrum more typical of a 100 Mpc box.
These four {\it halo catalogs} are summarized in Table~\ref{table: sims}.

Galaxies are identified initially as dark matter halos with ${{\delta\rho}/
{\rho}} > 30$ in 1-cell resolution elements (corresponding to about 4 cold
particles in a cell) which are local maxima in density. 
Halos with $M > 7 \times 10^{11} M_\odot$
were broken up to address overmerging (NKP96). 

We also tested filament and shape statistics on catalogs in which we identified
galaxy halos as cells with ${{\delta\rho}/
{\rho}} > 80$.  These catalogs
gave basic results which were quite similar to the halo catalogs
described above, with a slight increase in Poisson errors due to
fewer numbers of halos.  While  the ${{\delta\rho}/
{\rho}} > 30$ catalogs have too many halos to be associated with visible 
galaxies,
these catalogs still serve our purpose of testing whether these
statistics can quantify structure and discriminate between models in real space.
Comparisons with real data must be done using simulated redshift-space
catalogs.  

\subsection{The Sky Catalogs}

NKP94 and NKP96 describe the construction of the CfA1-like sky-projected
redshift catalogs from the simulations described in the previous section,
and the merged (to match simulation resolution) CfA1 catalog.
In order to distinguish these
catalogs which are designed to mimic many observational properties of the
CfA1 survey from the halo catalogs described above, we call the
CfA1-like sky-projected redshift catalogs the {\it sky catalogs}.
Several items in sky catalog construction which are 
of relevance to filament and shape statistics are: 

\begin{itemize}

\item Six view points were chosen from within the CHDM$_1$ 
and CHDM$_2$ simulations satisfying the conditions that
the local density in redshift space ($V < 750$ km s$^{-1}$) is within
a factor of 1.5 of the merged CfA1 galaxy density, and the closest Virgo-sized 
cluster is 20 Mpc away.  The CDM view points were required 
to be on the halos nearest to the CHDM$_1$ view point coordinates,
and thus the corresponding sky catalogs, like the halo catalogs, differ
only because of their underlying model physics and not cosmic variance.

\item To create a sky catalog of CfA1 size (12,000 km s$^{-1}$, 2.66 steradians), 
the periodic halo catalogs were stacked, then cut to form the CfA1 
survey geometry; hence structures appear typically $\sim 3-4$ times,
although distant galaxies are sampled sparsely.

\item Each sky catalog was cut to CfA1 numbers before fitting a Schecter luminosity 
function (after monotonically assigning Schecter luminosities to mass).
The scatter in Schecter function parameters among the six view points
is thus convolved into the statistics.

\end{itemize}

\subsection{The Effect of Halo Breakup}\label{sec: breakup}

The most massive halos in the simulation should generally have more than one individual 
galaxy associated with them (Katz \& White 1993, Gelb \& Bertschinger 1994).  
These ``overmerged'' halos were broken up as described
in NKP96 (it is the ``adopted method" set of catalogs that was used here).
Only 0.5\% of CHDM halos required breakup, raising the 
number of halos with ${\delta \rho / \rho} > 30$ by $\sim$16\%.
CDM1.5 and CDM1 catalogs had higher fractions of massive overmerged halos,
1.3\% and 1.7\% respectively, raising their breakup halo populations by 35\%
and 56\%, respectively.  
We expect the halo catalog results to be fairly insensitive to breakup since they
probe scales $\sim$3 Mpc and up, much greater than the radius over which
fragments are distributed, which is typically $\la 1$ Mpc.
Indeed we will show this to be the case in section 4.4.

Despite the larger scales investigated, sky catalogs will be more sensitive to breakup.
This is because breakup takes a single massive halo and fragments it
into many closely-distributed objects, 
many of which survive the magnitude limit.
When normalized to CfA1 number density, the net effect of breakup is to
weight the massive halos more strongly, giving the appearance on average of
moving galaxy halos into spherical groups (albeit with some ``finger of
God" elongation).  For a dense catalog, overdense
regions will be augmented at the expense of underdense regions, but
for sparse catalogs like CfA1, only the densest
clusters are augmented, at the expense of filamentary and planar structures. 
Hence halo breakup 
tends to systematically {\it reduce} the amount of filamentary and
planar structure measured in sky catalogs.  

\subsection{Cosmic Variance}

We will compare CHDM$_1$ to CDM1 and CDM1.5 to estimate the ability of
the statistics to discriminate between models, but by using
identical random number set initial conditions, cosmic variance is
explicitly removed.
Thus comparisons between these simulations reflect only differences
in the underlying physics of the models.
A proper measurement of the cosmic variance for these statistics
requires performing many simulations of each model varying the random sampling
of the initial power spectrum.  With limited computational resources,
we only have two such random samplings for a single model, 
{\it viz.,} CHDM$_1$ and CHDM$_2$.  
NKP94 and KNP96 estimate the high power
in CHDM$_1$/CDM1/CDM1.5 would be expected $\sim 10\%$ of the time,
translating to a $\sim 1.7\sigma$ deviation from norm, while CHDM$_2$
was found to be quite typical.  Thus CHDM$_1$ vs. CHDM$_2$ may be
taken as a crude estimate of $1\sigma$ cosmic variance.
However, a statistic which shows little difference between CHDM$_1$ and CHDM$_2$
does not necessarily have negligible cosmic variance, since
with only two realizations the possibility that the small deviation 
is merely a fortuitous coincidence for that statistic cannot be ruled out.
In the future, constrained realizations of the local universe should
bypass uncertainties from cosmic variance by constraining the poorly sampled 
large-scale waves in the simulation using redshift surveys (Primack 1995).

\section{Results for Halo Catalogs}

\subsection{Filament Statistics Applied to Halo Catalogs}

Figure~\ref{fig: dgfullstats} shows the results for 
filament statistics planarity $\bar\theta_P$, curvature $\bar\theta_C$, 
and torsion $\bar\theta_T$ vs. $R/\bar{d}$
applied to the halo catalogs catalogs after breakup.
The statistics were computed for each $R/\bar{d}$
from 1.2 to 2.5 in increments of 0.1.
To estimate errors in the halo catalogs, each statistic was computed
over a random subset of the catalog.  The subset was taken to be as
many halos as necessary to generate 500 link sequences.
Even for $R/\bar{d} = 1.2$, this never required more than 650 halos;
at high $R$, hardly a few percent of the halos generated sequences which 
did not meet the $N_{L,min} = 4$ criterion.  
The error bars shown in Figure~\ref{fig: dgfullstats} are $3\sigma$ resampling errors.
The catalog was then resampled 10 times to obtain an error estimate.  
Since there are more than 34,000 halos in each catalog, the data is
not oversampled.
At $R/\bar{d} = 1.2$, there were on average 5.6 links per sequence; this number rose
steadily until $R/\bar{d} \geq 1.6$, where sequences were was almost always 
terminated due to the $N_{L,max} = {100\;{\rm Mpc}/ \bar d} \approx 34$ criterion.  The
average number of halos within a sphere of radius $R$ around a 
given sequence point rose from
$\sim$10 at $R/\bar{d}=1.2$ roughly linearly to $\sim$50 at $R/\bar{d}=2.5$.

Figure~\ref{fig: dgfullstats} shows that all three statistics are 
generally higher for the CDM simulations as compared
with the CHDM simulations, indicating that 
CDM is less filamentary, has fewer sheet-like structures,
and has greater clumpiness than the CHDM simulations. 
These results are consistent with the notion that CDM possesses
more evolved structures with clumpier mass distributions, 
while the presence of neutrinos in CHDM models results in more
extended and less evolved structures
This notion is confirmed by the visualizations of BHNPK.  
Thus filament statistics provide quantitative differentiation
between large-scale structure seen in the halo catalogs.

Note that all the statistics tend to fall with increasing $R$.
This reflects the fact that as the ratio of ${R/L}$ increases, 
the greater overlap 
between adjacent spherical windows generates stronger correlations
between adjacent inertia tensors, thereby reducing the angle deviations 
between neighboring inertia ellipsoid axes.
There is an additional effect that is peculiar to catalogs possessing 
inherent filamentary structure:
Consider a link sequence tracing a path defined by 
points contained in a ``filamentary structure"
of radius $R_{cyl}$.  As we increase
${R/L}$ we see an increasingly more linear distribution of points in the
window, thus lowering the value of ${\bar\theta_C} \sim \frac{1}{2}
\arcsin(R_{cyl}/{R})$.  A similar argument holds for planarity
and torsion.  In reality the galaxy distribution
is more complex, but the basic result is that sampling large-scale structure 
gives ${\bar\theta_C}(R)$, ${\bar\theta_P(R)}$,
and ${\bar\theta_T(R)}$ 
falling at rates greater than in the Poisson case.

The large difference between simulations and the Poisson catalogs provides
a good indicator of how effectively structure
is identified by filament statistics.  
Link sequences identify and follow structure in a Poisson catalog
by detecting chance alignments of halos which masquerade as contiguous structure due to finite
numbers of halos in a given window.  As mentioned before, this
structure aliasing is primarily a low-galaxy-density phenomenon, and hence is 
most significant at low $R$, where all sequences barely exceed $N_{L,min}$,
and each window barely has $N_{P,min}$ halos.  In this situation the 
majority of sequences which qualify will be those lying along such
rare chance alignment of halos.  Increasing $N_{P,min}$ and $N_{L,min}$
reduces structure aliasing, but the corresponding reduction in qualifying
sequences increases shot noise significantly.  Instead, we simply
choose to be careful about our interpretations at low $R$.  For instance,
for $R/\bar{d} \leq 1.3$ the Poisson catalog statistics rise with $R$,
indicating that aliased structure is significant here.
Structure aliasing occurs in the models as well, but
is less apparent because halos are correlated, yielding
more halos surrounding a given point than in the Poisson case.
Nevertheless the reduced discrimination for $R/\bar{d} \leq 1.3$ is an
indication that aliased structure is of comparable strength 
to real structure at these scales.

At low and high $R$ values, filament statistics discriminate between
the CDM models with different biases, as shown in Figure~\ref{fig: dgfullstats}.
At $R/\bar{d} \leq 1.3$ CDM1.5 aliases structure more effectively than CDM1
since it is more diffuse (more Poisson-like),
while at larger scales ($\ga 7$ Mpc) the enhanced clustering of CDM1.5 
(see BHNPK) tends to
trap sequences in spherical clumps more effectively than CDM1, giving
higher values.
Identification of these effects over a $R$ of
1.0--2.5 (roughly 3.0--7.5 Mpc) indicates the high sensitivity of these
statistics to the presence of structure.  

The two CHDM simulation results are within $\sim 1\sigma$ of each other on
scales investigated.
Hence for these statistics, cosmic variance between CHDM$_1$ and CHDM$_2$ 
should be comparable to resampling errors in the halo catalogs.

\subsection{Shape Statistics Applied to Halo Catalogs}

We applied the BS, LV, and RA statistics to the halo catalogs after
breakup.  
To compare with filament statistics, we took 
ten sets of 1000 eligible halos each to compute resampling errors, 
varying the window radius $R$ from $1.2\bar{d}$ to $2.5\bar{d}$.
Eligible halos were 
those which had five or more other halos within the selected radius $R$,
analogous to the filament statistics computation.
BS points out that at least 12 halos are required within a window radius 
to avoid structure aliasing and reliably identify a planar
structure; however, for such a high value, few halos are eligible and
hence shot noise reduces the discriminatory power significantly.
Lowering this number to three increases the discriminatory power slightly,
but structure aliasing becomes more significant at small scales.
The results for $LV_{\sl quad}$ and $LV_{\sl plane}$ along with the $RA$
statistic applied to the halo catalogs 
after breakup are shown in Figure~\ref{fig: gfbulv}, while the results for the
$BS$ statistics are shown in Figure~\ref{fig: gfbubs}.
We omit $LV_{\sl line}$ for redundancy; it gives lower discrimination
than $LV_{\sl plane}$ but otherwise has very similar behavior.
The error bars shown are 3$\sigma$ resampling errors.

For $R/\bar{d} \leq 1.5$, the Poisson model is poorly discriminated
from the cosmological models for all statistics.  This is due to strong structure
aliasing in these small windows ($R \la 5$ Mpc),  making the statistics
clearly untrustworthy at these scales.
To avoid these spurious detections of structure we focus on the
regime $R/\bar{d} \geq 1.6$.  Recall that for filament statistics,
structure aliasing was problematic only for $R/\bar{d} \la 1.3$, so
filament statistics are able to discriminate true structure from aliased structure
at smaller scales.

For $R/\bar{d} \geq 1.6$, the results of the shape
statistics are consistent with BHNPK visualizations.  
CHDM models show higher filamentarity ($BS_{\sl prol}$ and $LV_{\sl quad}$)
and planarity ($BS_{\sl obl}$, $LV_{\sl plane}$, and $RA$) than CDM models,
while the sphericity measure $BS_{\sl sph}$ is higher for the CDM models.
However, CDM1.5 is generally closer to the CHDM models until $R/\bar{d}\ga 2.0$.
Comparing the values for $BS_{\sl prol}$ and $BS_{\sl obl}$
show that the halo distribution is more oblate than prolate, \ie
that large-scale structure in these models is dominated by 
sheets rather than filaments.  A comparison
of $LV_{\sl plane}$ and $LV_{\sl line}$ (not shown) yields the same conclusion.
Thus these shape statistics, like filament statistics, 
confirm and quantify the visually apparent differences between these models.

As $R/\bar{d}$ increases, the errors become smaller primarily due to more halos
being included within each window.  The statistical values also decrease
partly due to a reduction in structure aliasing, and partly because
larger windows tend to sample more spherical mass distributions.

All statistics appear to be fairly sensitive to the chosen
bias as well as the chosen set of initial conditions.  CDM1 is
well discriminated from CDM1.5 except in the region around
$R/\bar{d} \sim 2.5$ where their curves intersect; CDM1,
being the more evolved model, contains less filamentary and planar
structure at small scales.
For the LV statistics and $BS_{\sl sph}$, the cosmic variance estimated
from the difference between CHDM$_1$ and CHDM$_2$ begins to dominate
over resampling errors for $R/\bar{d} \ga 2.0$, while the $RA$ and
$BS_{\sl obl}$ shows $\ga 2\sigma$ cosmic variance at all scales.
Only for $BS_{\sl prol}$ is the cosmic variance always comparable to the 
resampling error for $R/\bar{d} \geq 1.6$.  

In comparison with filament statistics, the shape statistics give the
same conclusion regarding structure formation in the various models, but
appear to have more sensitivity
to cosmic variance (with the exception of $BS_{\sl prol}$),  and are 
more susceptible to structure aliasing at small scales.
The large difference between CDM1 and CDM1.5 indicates shape statistics
are more sensitive to the normalization of the cosmological model
than filament statistics; to some degree, this makes shape statistics
a complementary diagnostic.

\subsection{Measuring the Discriminatory Power}

We now introduce a set of metastatistics to compare statistics
and assess the effectiveness of our analysis.  These metastatistics
will allow a direct comparison of the discriminatory power and robustness
of filament statistics versus shape statistics.
Discrimination between models for a given statistic $\theta$ can be measured by 
the {\it signal strength} $S^\theta_{\rm res}$ between catalogs:

\begin{equation}
S^\theta_{\rm res}(1,2)={{| \theta_1-\theta_2 |}\over {\sqrt{\sigma_{\theta_1}^2+\sigma_{\theta_2}^2}}} 
\label{eq: dgsignal}
\end{equation}

\noindent
where $\theta_1$ and $\theta_2$ are values of statistic $\theta$ 
for catalogs 1 and 2, respectively,
and $\sigma_\theta$ is the resampling error for that statistic and catalog.  
The subscript ``res" denotes that the units of $S^\theta_{\rm res}$ are
$1\sigma$ resampling errors.
To compare CHDM to CDM at (roughly) COBE normalization while excluding
cosmic variance, we compare CHDM$_1$ to CDM1.
Figure~\ref{fig: dgsig}(a) shows $S^\theta_{\rm res}$(CDM1,CHDM$_1$) for the halo catalogs
for $\theta = \{\bar\theta_P, \bar\theta_C, \bar\theta_T\}$.
computation) 
For $R \geq 1.4$, where structure aliasing is unimportant, 
planarity shows the highest signal, then curvature then torsion, 
with all statistics showing $S^\theta_{\rm res}$(CDM1,CHDM$_1$)$\ga 4\sigma$.
Thus filament statistics are fairly discriminatory for the halo catalogs;
their robustness against halo breakup will be formally investigated in 
\S \ref{sec: robustness}.
Cosmic variance is not expected to dramatically degrade the discrimination
as the differences between CHDM$_1$ and CHDM$_2$ are comparable to
the resampling errors.

The signal strengths for the LV, BS, and RA statistics applied to the
halo catalogs after breakup are shown in Figure~\ref{fig: dgsig}b (bottom panel).
For most statistics, discrimination is
better than $5\sigma$ for $R/\bar{d} \geq 1.6$.  The significant
exception is $BS_{\sl prol}$, with barely $\sim 2-3\sigma$ discrimination
for $R/\bar{d} \geq 1.8$.  However, recall that $BS_{\sl prol}$
was also the only statistic whose cosmic variance was small
compared with resampling errors.  With cosmic variance taken into account,
$BS_{\sl prol}$ appears to be quite comparable in discriminatory power
to the other statistics.  In general, both filament and shape statistics
appear to be easily discriminate structure formation in CDM models
versus CHDM models.  Cosmic variance, though, appears to be more of a concern
for the shape statistics, with the exception of $BS_{\sl prol}$.

\subsection{Robustness Against Halo Breakup}\label{sec: robustness}

The identification of galaxies in simulations represents a major
uncertainty in this type of analysis.  Given that the simulations
are dissipationless, it is not possible to directly identify clumps of
baryons which would be expected to form galaxies.  Instead, assumptions
must be made regarding how the baryonic matter traces the dark
matter.  In addition, because of the limited resolution of the
simulations, a single clump of dark matter may contain several galaxies
(often referred to as the ``overmerging problem"), and must be
broken up to obtain a true sample of galaxies.  The detailed
assumptions made in this procedure (described in NKP96)
are somewhat {\it ad hoc}, so it is important to somehow quantify
the uncertainty introduced by our lack of knowledge.

To do this we compare the statistical values for the catalogs before
breakup and after breakup.  The effects of breakup on these statistics
are expected to be as follows: Because single halos
are broken into several spherically-distributed halos, the tendency
will be to decrease the detection of planar and filamentary structure,
and increase the detection of spherical structure.  This is in fact
what is seen.  Since it is clear some sort of halo breakup must be done, and
that halo breakup has a monotonic effect on the statistics, a
comparison of catalogs before and after breakup should yield a
somewhat conservative estimate of the uncertainty introduced by
this procedure, unless the breakup scheme used here produces far
too few fragments.

We first measure the {\it halo identification uncertainty factor}, given by
\begin{equation}
F_{\rm id}(\theta)=\left\langle{{| \theta{\rm bu}-\theta{\rm nobu} |}\over
{\sqrt({\sigma_{\theta{\rm bu}}}^2+{\sigma_{\theta{\rm nobu}}}^2)}}\right\rangle_{{\rm cats}}
\end{equation}

\noindent
where $\theta$ represents the value of the statistic in question, $bu$ and $nobu$
refer to breakup and no-breakup catalogs,
and $\sigma_{\theta}$ represents the resampling error for statistic $\theta$.
We then combine this error with resampling errors to obtain the {\it combined
signal strength} metastatistic, presented as a function of 
radius $R$:
\begin{equation}
S^\theta{\rm res+id}(1,2)={{S^\theta{\rm res}(1,2)}\over {MAX[1.0,F_{\rm id}(\theta)]}}
\label{eq: dgrob}
\end{equation}
\noindent 

For the filament statistics applied to the halo catalogs, we compute 
$S^\theta_{\rm res+id}({\rm CHDM}_1,{\rm CDM}_1)$ for each statistic 
for $R = 1.2-2.5$.  The results are plotted in Figure~\ref{fig: dgrobful}(a)
(top panel).  
Torsion show no degradation of signal, as $F_{\rm id}(\theta_T) < 0.4$ for
all values of $R$; this statistic is highly robust against halo identification
uncertainty.  Curvature, conversely, shows some degradation, as it typically
has $F_{\rm id}(\theta_C) \sim 1.7$.  Planarity, which showed the highest
signal strength, is by far the least robust, with $F_{\rm id}(\theta_C) \sim 2$
and as high as 2.5 at some values of $R$.  Comparing with
Figure~\ref{fig: dgsig}(a), torsion now appears marginally to be 
the best filament statistic, showing typically $4\sigma$ robust
discrimination between models, while curvature and planarity have 
$S^\theta_{\rm res+id}({\rm CHDM}_1,{\rm CDM}_1) \sim 3$.  
Recall from Figure~\ref{fig: dgfullstats} that all statistics show CHDM$_1$ and
CHDM$_2$ being $\la 1\sigma$ apart, so these conclusions should not be
dramatically affected by cosmic variance.

Figure~\ref{fig: dgrobful}b (bottom panel) 
shows the combined signal strengths computed shape statistics applied 
to the halo catalogs.   For all the statistics except $RA$, 
the results are generally insensitive to breakup.
For $RA$, $F_{\rm id} \sim 1.5$ typically, but it still leaves $RA$
with comparable discriminatory power as other shape statistics.
While no shape statistic is clearly optimal by this measure,
$BS_{\sl obl}$, $BS_{\sl sph}$, and $LV_{\sl plane}$ appear to show the
strongest discrimination; however cosmic variance is a concern for
those statistics.  $BS_{\sl prol}$, which had low cosmic variance,
is not very discriminatory.
Overall, there are statistics from each category showing over $4\sigma$
discrimination which is robust against variations in the galaxy identification
scheme.

\section{Results For Sky Catalogs}

\subsection{The Effect of Redshift Distortion}\label{sec: reddist}

Distortion of structure due to peculiar motions of individual
galaxies (\eg fingers of God) could in principle significantly
degrade the ability of all of these statistics to quantify true structure.
In our case, since CDM contains higher peculiar velocities, one
might expect stronger fingers of God which could mimic the
true filamentarity contained in CHDM and thereby work against
the discriminatory power of these statistics.  
This does not turn out to be the case, however, because
fingers of God are elongated only in the line-of-sight direction {\bf \^r} 
in redshift space,
whereas in general a structure in real space will not be aligned with {\bf \^r}.
Thus redshift distortion tends to smear out and hence {\it decrease} the amount of
structure detected in these simulations, slightly more so in models with
more redshift distortion.
For filament statistics, a further effect of redshift space distortion is 
to misguide sequences and
increase the angle deviation of a sequence passing through a cluster.
The net result is that redshift distortion
does not significantly undermine the
ability of any of these statistics to distinguish between cosmological models,
and in some cases slightly enhances the discrimination.

To test the effect of redshift distortion we adopt the strategy of
applying the statistics to mock redshift catalogs constructed to exaggerate the
distortion due to peculiar velocities.
We first cut the halo catalogs at a high density
threshold, roughly mimicking CfA1 sparseness.  Then for each halo we compute the
line-of-sight
velocity ${\rm v}_{\rm los}$ with respect to an observer at one corner of the
simulation volume.  We then multiply this velocity by
$F_V$, the {\it velocity scaling factor}, and shift the halo
position along the line of sight by
\begin{equation}
\Delta {\bf x} = F_V {\rm v}_{\rm los} \Delta\hat{\bf r}/H_0
\end{equation}
where $\Delta\hat{\bf r}$ is the direction from the box center to the given
halo.  Thus $F_V = 0$ corresponds to real space, $F_V = 1$ corresponds
to redshift space, while higher $F_V$ yields an exaggerated shift
from which we can gauge the sensitivity of the statistics
to redshift distortion.  We choose $R/\bar{d} = 1.5$ when applying the test to
filament statistics, and $R/\bar{d} = 1.8$ when testing 
the other shape statistics.

Figure~\ref{fig: reddistfs}(a) (left side) shows
the effect of the transformation from real to redshift space upon filament
statistics for halo catalogs with a density threshold of 
${\delta \rho}/{\rho} > 80$, while Figure~\ref{fig: reddistfs}(b) 
(right side) 
shows the equivalent results for ${\delta \rho}/{\rho} > 120$ catalogs.
At $F_V = 1$, we see a definite increase in the discrimination
between the models as compared to  $F_V = 0$, most notably for
the curvature and torsion statistics.  This effect is more
pronounced in the sparser catalogs.  The statistical values increase,
indicating structure is being smeared out by redshift distortion.
The trend continues to higher $F_V$, with no dramatic dropoff in
the discriminatory power of filament statistics.
This test was also run on the pre-breakup versions of the same catalogs,
and it was found that the interpretations are virtually independent of 
breakup in both real and redshift space.  

Figure~\ref{fig: reddistshp} shows
the results of the redshift distortion test for three selected shape statistics
(the others show similar behavior).  As with filament statistics, 
less structure is detected when redshift distortion is included.
Unlike filament statistics, however, there is no apparent increase in the 
disriminatory power of these statistics at $F_V = 1$.  Also, 
exaggerated redshift distortion ($F_V \geq 2$) has little further
effect on the statistics.  Again, these results are fairly insensitive to 
breakup and catalog density.

In summary, the primary effect of redshift distortion is to
decrease strength of structure, which if anything will work to amplify the 
discrimination between the models considered.  
Filament statistics, which emphasize regions
of higher density, are more affected by
redshift distortion than the randomly-sampled shape statistics, and
we see greater amplification of discrimination between models.

\subsection{Filament Statistics Applied to Sky Catalogs}

Figure~\ref{fig: dgskystats}
shows the results of filament statistics applied to the sky catalogs
after halo breakup.
Every galaxy in each sky catalog was tried
as a possible sequence starting point.  For each catalog, at $R = 1.2$, 
around 800 of the $\approx$2360 galaxies typically generated sequences with
number of links exceeding $N_{L,min}=4$.  This number rose roughly linearly until
$R=2.5$, where $\sim$2200 galaxies qualified, on average,
in each catalog.  There were systematic differences between the catalogs
as well, with CHDM$_2$ showing the largest number of accepted sequences,
about $5-10\%$ more than the CDM models.  CHDM$_1$ showed the
lowest number, consistently slightly below the CDM models.
At $R = 1.2$, there were on average about 6 links per sequence;
this number rose fairly linearly with $R$, such that at $R=2.5$, there 
were around 20 links per sequence.  The average number of galaxies within
a sphere of radius $R$ around a given sequence point rose from 8--10 
at $R=1.2$ roughly linearly to 25--30 at $R=2.5$.  

The error estimate for each statistic in sky catalogs was
determined from {\it sky variance}, by computing the statistic at
each of six vantage points, and getting an average value and standard
deviation for that statistic.
The error bars shown in Figure~\ref{fig: dgskystats} are $1\sigma$ sky variance errors.
Since our box is relatively small, different viewpoints are still
seeing many of the same structures, although with differing depth.
Sky variance is therefore expected to underestimate true cosmic
variance, perhaps significantly.

Figure~\ref{fig: dgskystats} shows that both
CHDM models still show more structure than either CDM model,
consistent with our intuitive picture of structure formation in these models,
and all models are fairly well discriminated from the Poisson catalog.
However, CHDM$_1$ shows significantly more structure than CHDM$_2$,
by up to $\sim 2\sigma$ for the torsion and curvature statistics,
indicating that that sky variance is an inadequate estimate of cosmic variance.
The extra large scale power in CHDM$_1$, accentuated by
the artificial replication of structure in the construction
of the sky catalogs at 100 to $100\sqrt{3}$~Mpc intervals,
produces more large-scale structure in CHDM$_1$ than in CHDM$_2$.
This is more apparent in sky catalogs than in the halo catalogs
since the scales investigated are much larger, with $\bar{d}\sim 10$ Mpc
even in the region where the sky catalogs are complete.

To test sensitivity to shot noise and catalog boundary effects, 
filament statistics were applied to (nearly) full-sky versions of
the CfA1-like sky catalogs with a zone of avoidance $|b| \leq 10^\circ$ about
each viewpoint, covering 10.384~sr instead of 2.66~sr and containing
about four times as many galaxies ($\approx 9200$).  
Since the 2.66~sr catalogs and the 10.384~sr catalogs are derived 
from the same simulation data set, we are
still sampling from the same distribution of cluster sizes and shapes.
The resulting signal strength increased by a factor of $\sim 2$ (for $R/\bar{d} \geq 1.3$)
as expected if the errors are dominated by shot noise.
The degradation of the signal from the halo catalogs to the sky catalogs
is thus primarily due to sparseness.  
For a survey such as the Optical Redshift Survey (Santiago 
\etal 1995,1996) which covers 8.09 sr at CfA1 depth, 
we expect to see well over $3\sigma$ discrimination between models,
excluding cosmic variance.

We quantify boundary effects by comparing statistical values for the 
2.66 sr catalogs vs. the 10.384, and find that for $R/\bar{d} \ga 2.0$, 
the CfA1-like catalogs show
significantly higher values (comparable to sky variance) 
than the 10.384~sr catalogs, indicating that 
the entire catalog volume was contributing as a single radial filamentary 
structure.
This was also evident from visualizations of the link sequences, as
at large $R$ the sequences were preferentially radially directed.
Visualization also showed that link sequences were distributed throughout
the sky catalog volume, with very few lying in the foreground, $r\la 20$ Mpc.  
Recall that
$\bar d (r)$ is small at low $r$, and the Virgo Cluster, being nearby,
contributes hardly any sequences even though it gives a large
finger of God.  At small $R$, sequences tended to be shorter and
terminate within the catalog volume, while at large $R$ they tended
to terminate once they exceed the catalog boundary and find no
nearby galaxies.  

The statistics were also applied to 80 Mpc volume-limited versions of the sky
catalogs, with typically 400-500 galaxies in each.
The statistics showed very large shot-noise
scatter, and gave no significant discrimination between models.
Volume limiting certainly yields more interpretable statistics, 
but for CfA1 and our similar-size simulation sky catalogs, 
there are simply too few galaxies.

\subsection{Shape Statistics Applied to Sky Catalogs}

In Figure~\ref{fig: gcbulv} we present the (selected) $LV$ and $RA$ statistics
and in Figure~\ref{fig: gcbubs} the $BS$ statistics,
applied to the sky catalogs after halo breakup.  
Also plotted as solid lines are the results for the CfA1 catalog.   
Each statistic was computed around every galaxy in the sample.
The errors are increased greatly
over the halo catalog case because of the sparseness of these
CfA1-like catalogs; the error bars shown are $1\sigma$ sky variance errors.

The interpretation of statistical values again
confirms the intuitive picture of structure formation in these
models.  CHDM models show greater filamentarity and planarity 
and less sphericity than CDM models for $R/\bar{d} \ga 1.5$.
As with the halo catalogs, the galaxy distribution shows stronger 
planarity than filamentarity, with $BS_{\sl obl}$ typically
twice $BS_{\sl prol}$ at any given $R$.
Also, the CDM models show different trends
versus $R$ analogously to the halo catalogs, with CDM1.5
dropping faster versus $R/\bar{d}$ than CDM1.

For the LV and RA statistics applied to the sky catalogs (Figure~\ref{fig: gcbulv}),
structure aliasing is a significant concern.  
The Poisson catalog is not discriminated from the models until
the scales are quite large, $R/\bar{d} \sim 1.8$ for $LV_{\sl quad}$
and $RA$.  The higher order LV statistics (represented in
Figure~\ref{fig: gcbulv} by $LV_{\sl plane}$; $LV_{\sl line}$
and $LV_{\sl flat}$ show similar behavior) are never well discriminated
from the Poisson catalog.  
As with the halo catalogs, filament statistics do a better job avoiding aliased
structure at small scales.
The CHDM models are well separated, indicating
that cosmic variance dominates over sky variance for these
statistics.  In all, the LV and RA statistics appear less able to
reliably quantify structure than filament statistics
in a catalog as sparse as CfA1.

The BS statistics (Figure~\ref{fig: gcbubs}) do not have quite as much
difficulty distinguishing a Poisson catalog from the cosmological
models as the LV and RA statistics, although they still
cannot discriminate for $R/\bar{d} \leq 1.5$.
$BS_{\sl prol}$, just as in the halo catalog case, shows
remarkable little cosmic variance for $R/\bar{d} \ga 1.8$,
although with only two realizations of CHDM
the possibility that this is merely a fortuitous coincidence 
cannot be ruled out.
For $BS_{\sl obl}$ and $BS_{\sl sph}$, cosmic variance 
is again a significant source of uncertainty, with CHDM$_1$
generally closer to the CDM models than CHDM$_2$.

We also applied these statistics to the full-sky versions (10.384~sr) of 
the sky catalogs.  The discriminatory power of the
best statistics increased only to $\sim 3\sigma$, showing that
these statistics are not completely dominated by 
Poisson noise, and are more affected by halo identification uncertainty
than filament statistics.
As with filament statistics, boundary effects become significant 
for $R/\bar{d} \ga 2.0$.

\subsection{Robustness and Discrimination Between Models}

In Figure~\ref{fig: dgrobcfa} we present the combined signal strength
$S^\theta_{\rm sv+id}$(CDM1,CHDM$_1$) for all the statistics applied to
the sky catalogs.  
The subscript ``sv" signifies that we are including sky variance errors.
The results before breakup are not shown, but for most statistics,
the sparseness of the CfA1 catalog generates
sky variance errors which dominate over
halo identification uncertainty, so $F_{\rm id}< 1$ at nearly all $R$.  
The exception is the $RA$ statistic, which had $F_{\rm id} \sim 1.5$ typically.
As described in section~\ref{sec: breakup}
the catalogs before breakup show slightly more structure than after breakup.
It turns out that for filament statistics, this
represents a $\la 1^\circ$ increase in each statistic for the 
sky catalogs, which is generally less than sky variance errors.  
There is little qualitative difference in
$S^\theta_{\rm sv+id}$(CDM1,CHDM$_1$) for no-breakup sky catalogs.

Figure~\ref{fig: dgrobcfa}(a) shows that for filament statistics,
discrimination between CHDM$_1$ and CDM1 is strongest in torsion 
($\sim 2.5\sigma$) and curvature ($\sim 1.5-2\sigma$), 
while planarity shows no significant discrimination between CHDM and CDM.  
Planarity is weaker
because it is not as significantly amplified by redshift distortion
as curvature and torsion, as was described in
\S \ref{sec: reddist} (see Figure~\ref{fig: reddistfs}(b)).
While promising, these levels of discrimination are comparable to our crudely
estimated cosmic variance.

The signal strengths $S_{\rm sv+id}$(CHDM$_1$,CDM1) for the shape statistics
applied to the sky catalogs after breakup
are shown in Figure~\ref{fig: dgrobcfa}b (bottom panel). 
Greatest discrimination is seen for $BS_{\sl prol}$,
at a modest $\sim 1.5-2\sigma$ level for $1.7 \la R/\bar{d} \la 2.2$.
The $LV_{\sl line}$ statistic shows some apparent discrimination at 
$R/\bar{d} \sim 1.5$, but recall for this $R$ this statistic does
not discriminate a Poisson catalog from the cosmological models.
None of the other statistics show significant discrimination between
these models.

Overall, we conclude that for the CfA1-like sky catalogs, the
best filament statistic is torsion, which clearly has the greatest
discriminatory power of any statistic with the caveat that it may be 
significantly degraded by cosmic variance.
Of the shape statistics, the Babul \& Starkman prolateness measure
$BS_{\sl prol}$ gives the most discrimination between 
models when applied to a CfA1-like data set,
showing some discriminatory power (up to $2\sigma$) and good robustness 
against both halo identification uncertainty and cosmic variance.  
The LV and RA shape statistics show little discriminatory power
between models or even from a Poisson catalog in such a sparse survey.
Testing on 10.384~sr versions of the sky catalogs shows that 
all these statistics are hampered mostly by shot noise,
thus a larger data set is required to properly discriminate
between these models.  The Optical Redshift Survey (Santiago \etal 1995, 1996)
which has CfA1 depth but 8.09 sr sky coverage, will be very useful
for this purpose.

\subsection{Comparing Models vs. CfA1 Data}

Using these statistics we can compare the sky catalog results directly 
to CfA1 data.  For filament statistics shown in Figure~\ref{fig: dgskystats},
the CfA1 catalog follows the CDM models more closely than the CHDM models.
However, given the uncertainty in halo identification ($\sim 1^\circ$) and
cosmic variance, it is difficult to conclusively state
which model agrees best with filament statistics based on the CfA1 data set.

The various shape statistics presented in Figures~\ref{fig: gcbulv} and
\ref{fig: gcbubs} likewise
show that no single model of those considered here 
is completely consistent with CfA1 data.
The LV and RA statistics show best agreement with CHDM$_1$,
but these statistics have large cosmic variance.
$BS_{\sl obl}$ and $BS_{\sl sph}$ show best agreement with the CHDM models, 
while $BS_{\sl prol}$ shows best agreement with the CDM models.
With at best $2\sigma$ discrimination combined with 
the uncertainties in our estimates of cosmic variance and
halo identification robustness, 
we again cannot favor or rule out any models based on these shape
statistics applied to the CfA1 redshift survey.

\section{Conclusions}

In this paper we present filament statistics, a new set of 
statistics for quantifying filamentarity and
planarity in large-scale structure.
We compare these statistics to the shape statistics of
Babul \& Starkman (1992), Luo \& Vishniac (1995),
and Robinson \& Albrecht (1996)
by introducing metastatistics which quantify the discriminatory
power and robustness of each statistic.
We find that when applied to the halo catalogs, 
most of the statistics considered are sensitive and robust 
diagnostics of large scale structure that effectively discriminate simulations 
of CDM models from simulations of CHDM models, with 
robust discrimination of $\ga 4\sigma$ 
between CDM and CHDM models.  Cosmic variance is low for the
filament statistics, but more of a concern for all the shape
statistics except perhaps $BS_{\sl prol}$.
The signal-to-noise ratio
between any model and the Poisson catalog is very large for
all $R \geq 1.5\bar{d}$, where $R$ is the window radius and $\bar{d}$
is the mean intergalaxy spacing.
Finally, all statistics show that CHDM contains more sheet-like and
filamentary structures than CDM, consistent with intuitive expectations
as well as visualizations done by BHNPK.

Comparison with redshift survey data must be done in redshift space with the
appropriate survey geometry.  We compare CDM and CHDM models to CfA1 data
by utilizing a sample of CfA1-like redshift catalogs constructed
from each of the simulations,
and comparing these ``sky catalogs" directly to the CfA1 survey.
When one views the statistics' results for sky catalogs,
it is unclear which statistic provides the most discrimination between models.
Filament statistics tend to show better robust discrimination than
the shape statistics, but cosmic variance is a concern.
Comparing models to CfA1, we find the filament statistics
show, at face value, that the CDM simulations provide the best fit to CfA1 data.  
On the other hand, for most shape statistics the CHDM models
appear to be a better fit.  In all cases
the discrimination is poor, and significantly weakened by
uncertainties in halo identification as well as cosmic variance.
A proper comparison of statistics and of models versus
redshift survey data must await larger data sets.


In a broad context, we view filament statistics as illustrative
of a new methodology for constructing statistics to analyze spatial data.
We utilize inertia tensors to characterize the local mass distribution,
similarly to the LV, BS, and RA statistics.
But rather than deriving combinations of
tensor moments to quantify structure,
filament statistics use link sequences to generate new data samples which
amplify properties of interest in the underlying data set.  
The link sequence approach was conceived of as an intuitive means
of simplifying the complex topology of the galaxy point set while
enhancing the sense of approximate connectivity of
its large-scale isodensity surfaces (which the eye might recognize as 
``filamentarity'').  Since the link sequences are 
{\it guided} by the distribution of galaxies, not {\it bound} by it
(as in Delaunay or Voronoi tessellations, see e.g. van de Weygaert 1991,
or minimal spanning trees, see e.g. Pearson \& Coles 1995),
they are more likely to be robust against variations in the galaxy 
locations and halo breakup, although as we have seen, robustness
against galaxy identification in magnitude-limited mock redshift
catalogs is a trickier issue.  
Another approach for using link sequences is to apply
shape statistics like those of LV, BS, and RA
to the newly created data sample, producing statistics which may be
more discriminatory than any presented here.  We plan to investigate this
possibility in the future.

The success of these statistics for the halo catalogs indicates that
larger, denser redshift surveys coupled with larger simulations
will provide a significant
increase in the robustness and discriminatory power of these statistics
versus real survey data.  
A proliferation of such large redshift surveys is already underway.
On the simulations front, good
progress is being made in scaling up the size and resolution of cosmological
simulations, as well as in constructing constrained
realizations of the local universe (Primack 1995) by which one may
avoid uncertainties of cosmic variance.  Thus we soon hope to have
a suite of significantly larger simulations of currently favored
models which we can compare to these large redshift surveys.
Finally, there is interesting work being done in more realistically
handling the overmerging problem by combining approximations to
hydrodynamics with Press-Schecter type formalisms to accurately
model the numbers of galaxies near the resolution limit of the
simulations (Kauffman, Nusser \& Steinmetz 1995;
Somerville \etal 1996, in preparation).  In the coming years 
we hope to establish
these statistics which quantify the shapes of large-scale structure 
as significant constraints on cosmological models of structure formation.

\section*{Acknowledgements}
We thank Ethan Vishniac for helpful discussions.
We acknowledge grants of computer resources by IBM, NCSA, SCIPP,
UCO/Lick Observatory, and UCSC Computer Engineering.
RD acknowledges support from Lars Hernquist under NSF grant ASC 93-18185.
AK, JRP, and DH acknowledge support from NSF grants and
DH also acknowledges support from a DOE grant.
The simulations were run on the Convex C-3880
at the NCSA, Champaign-Urbana, IL.  

\def\apj{ApJ}
\def\apjs{ApJ Supp}
\def\apjl{ApJ Lett}
\def\mnras{MNRAS}
\def\aa{Astron. \& Astrophys.}
\def\aj{AJ}

\onecolumn


\bigskip

\begin{table}
\caption{Halo catalogs (KNP96,NKP96)}
\label{table: sims}
\begin{tabular} {@{}lcccccc}
{Model} & {$\Omega_c/\Omega_\nu/\Omega_b$} &  {Bias} & {Q$_{\rm rms}$($\mu$K)} & {Init.Cond.} & {No. of Gals.} & {$\bar d$ (Mpc)} \\
CDM1 & $1.0/0/0$ & $b=1.0$ & 12.8 & Set 1 & 58,121(37,164) & 2.58(3.00) \\ 
CDM1.5 & $1.0/0/0$ & $b=1.5$ & 8.5 & Set 1 & 61,690(45,592) & 2.53(2.80) \\ 
CHDM$_1$ & $0.6/0.3/0.1$ & $b=1.5$ & 17.0 & Set 1 & 34,000(29,151) & 3.09(3.25) \\ 
CHDM$_2$ & $0.6/0.3/0.1$ & $b=1.5$ & 17.0 & Set 2 & 34,554(29,765) & 3.07(3.23) \\ 
\end{tabular}
\medskip
{\\ The number of galaxies and mean interparticle spacing $\bar d$ computed 
before halo breakup are indicated in parentheses.}
\end{table}


%
%
%

\twocolumn

\bigskip
\begin{table}
\caption{Shape statistics test cases: Line, Plane, and Sphere}
\label{table: test}
\begin{tabular} {@{}lccc}
{Stat} & {Line} & {Plane} & {Sphere} \\
$LV_{\sl quad}$ & 0.976/0.989 (1)& 0.269/0.254 (0.25)& 0.012/0.006 (0)\\ 
$LV_{\sl line}$ & 0.953/0.978 (1)& 0.035/0.008 (0)& 0.004/0.002 (0)\\
$LV_{\sl plane}$ & 0.000/0.000 (0)& 0.906/0.978 (1)& 0.029/0.014 (0)\\
$LV_{\sl flat}$ & 0.953/0.978 (1)& 0.959/0.989 (1)& 0.033/0.015 (0)\\
$BS_{\sl prol}$ & 1.000/1.000 (1)& 0.016/0.002 (0)& 0.004/0.002 (0)\\
$BS_{\sl obl}$ & 0.000/0.000 (0)& 0.959/0.991 (1)& 0.012/0.001 (0)\\
$BS_{\sl sph}$ & 0.000/0.000 (0)& 0.000/0.000 (0)& 0.909/0.977 (1)\\
$RA$ & 0.000/0.000 (0)& 0.659/0.743 (1)& 0.168/0.063 (0)\\
\end{tabular}
\medskip
{\\ Values of individual statistics for three test case random distributions.
The first value is for $R=5$ Mpc, the second is for $R=10$ Mpc, and
the value in paranthesis is the analytical value for that distribution.}
\end{table}

\onecolumn


\section*{Captions}

\bigskip
\noindent Figure~\ref{fig: dg_flowchart_filament}.
Link sequence generation computational flowchart.  
$R$ is taken in units of the mean intergalactic spacing $\bar d$.  
For galaxies in redshift space, $\bar d$ is a function of the Hubble distance
$r = v/H_0$, where $v$ is the radial velocity of the galaxy.
From the initial galaxy, sequences are propagated in both (opposing) directions
along the major axis until termination; if the combined number of links
is 4 or more, the entire (combined) sequence ``qualifies" for computation; else
it is discarded.

\begin{figure}
\plotone{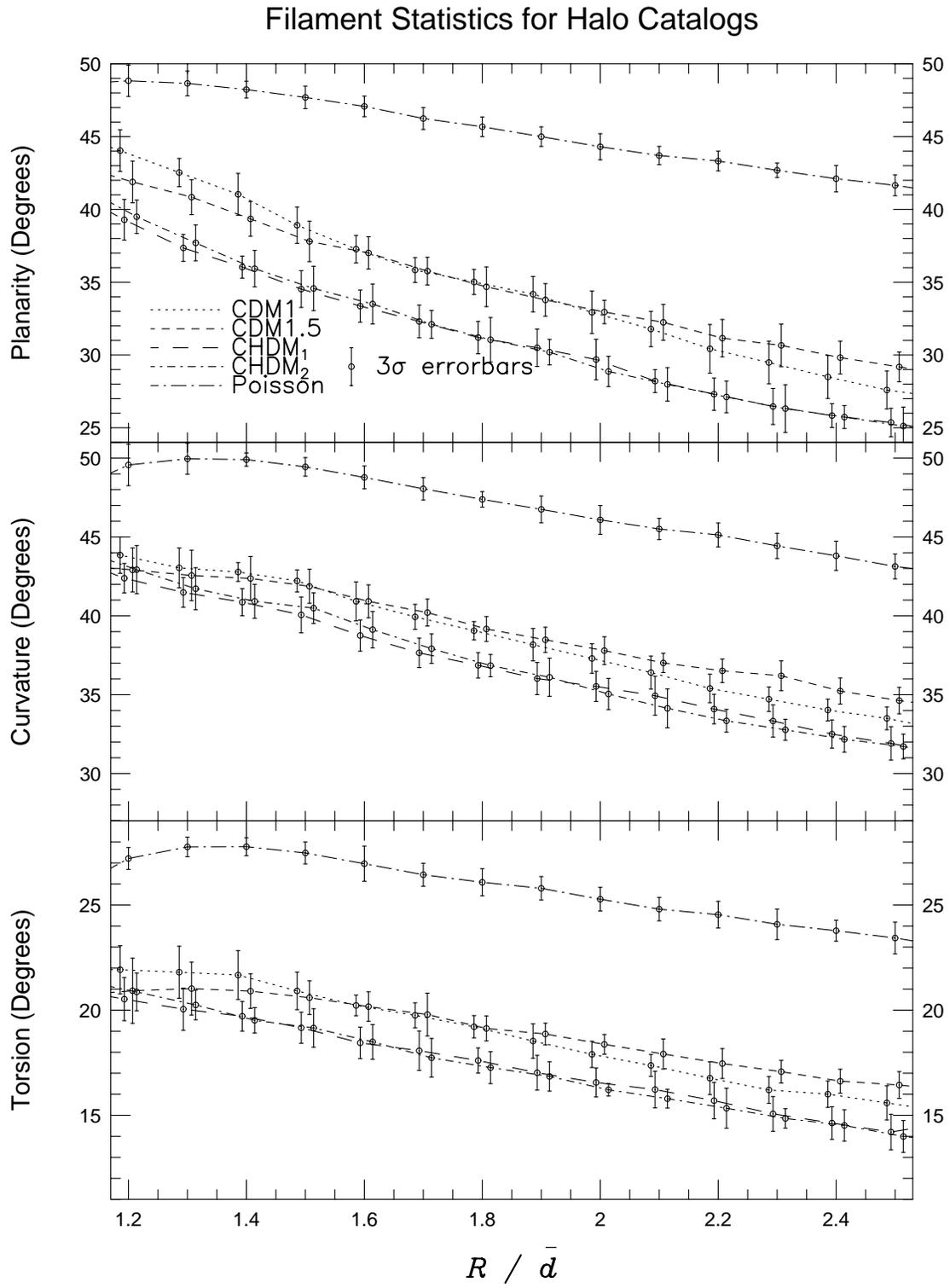}
\caption{Halo Catalogs Filament Stats}
\label{fig: dgfullstats}
\end{figure}

\bigskip
\noindent Figure~\ref{fig: dgfullstats}.
{Filament statistics (planarity $\bar\theta_P$, curvature $\bar\theta_C$, and torsion $\bar\theta_T$) 
for the {\it halo catalogs} versus $R/\bar{d}$, with $L=\bar d$.
Error bars shown are $3\sigma$ resampling errors.
The statistics show the CHDM models having more structure than
the CDM models, by well over $4\sigma$ at most $R$.
Cosmic variance estimated by the difference between
CHDM$_1$ and CHDM$_2$ is generally comparable to resampling error.
The Poisson catalog is well discriminated from any model
for $R/\bar{d}\geq 1.3$.
Note: Values for different models are slightly offset in $R$ to improve
visibility.}

\begin{figure}
\plotone{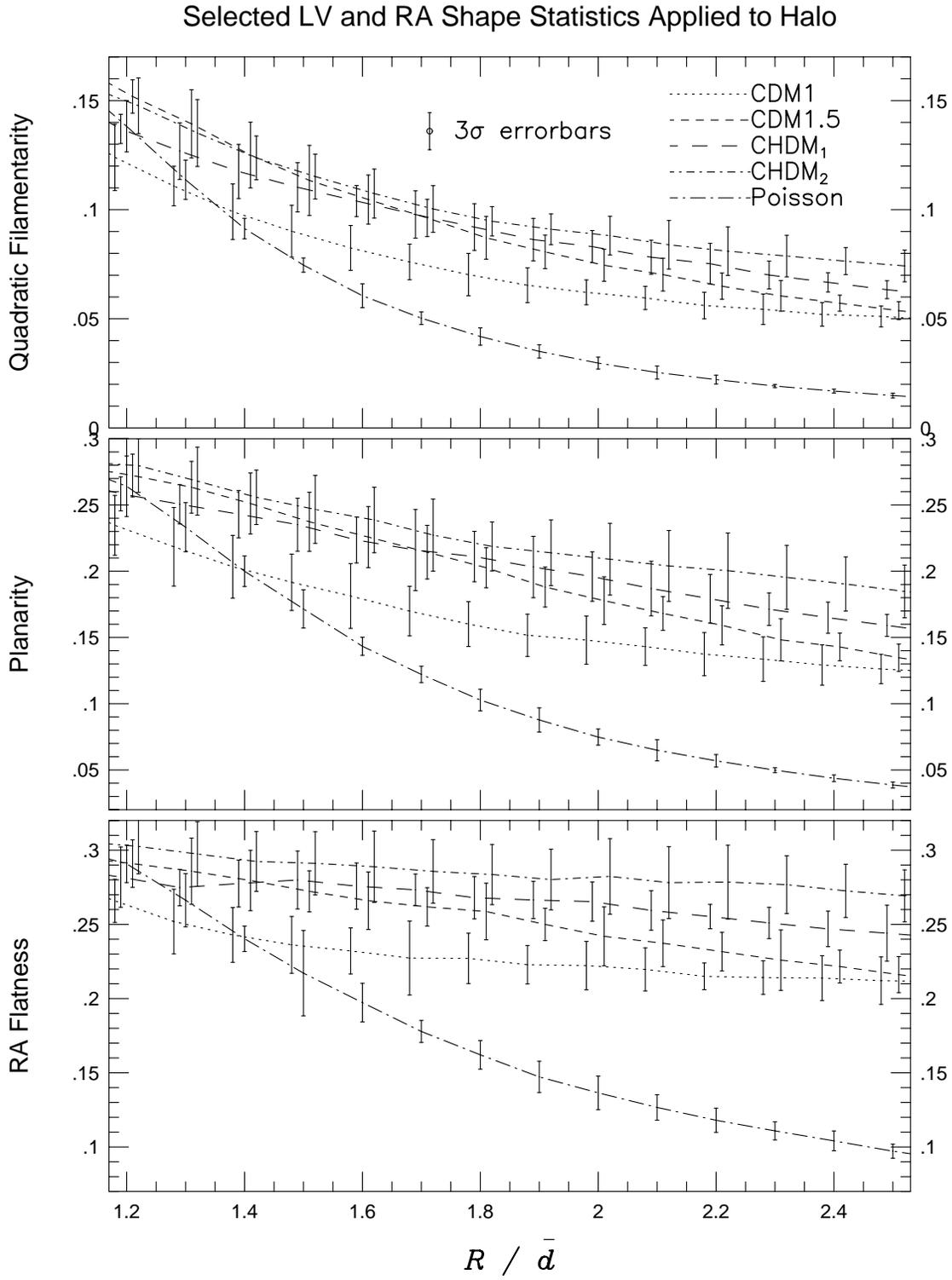}
\caption{Halo Catalogs LV \& RA Stats}
\label{fig: gfbulv}
\end{figure}

\bigskip
\noindent Figure~\ref{fig: gfbulv}:
Results for selected Luo \& Vishniac (1995) statistics 
and the Robinson \& Albrecht (1996) statistic applied to the halo catalogs
after halo breakup.  
Error bars shown are $3\sigma$ resampling errors.
The cosmological models are well discriminated from the Poisson model for
$R/\bar{d} \geq 1.6$.  The statistics are sensitive to 
the cosmological model as well as to the normalization
(\ie the bias factor), with CDM1 and CDM1.5 showing markedly different
trends with $R$.
Cosmic variance seems to be significant for all these statistics, indicating
that resampling errors may not be an appropriate measure of total variance.

\begin{figure}
\plotone{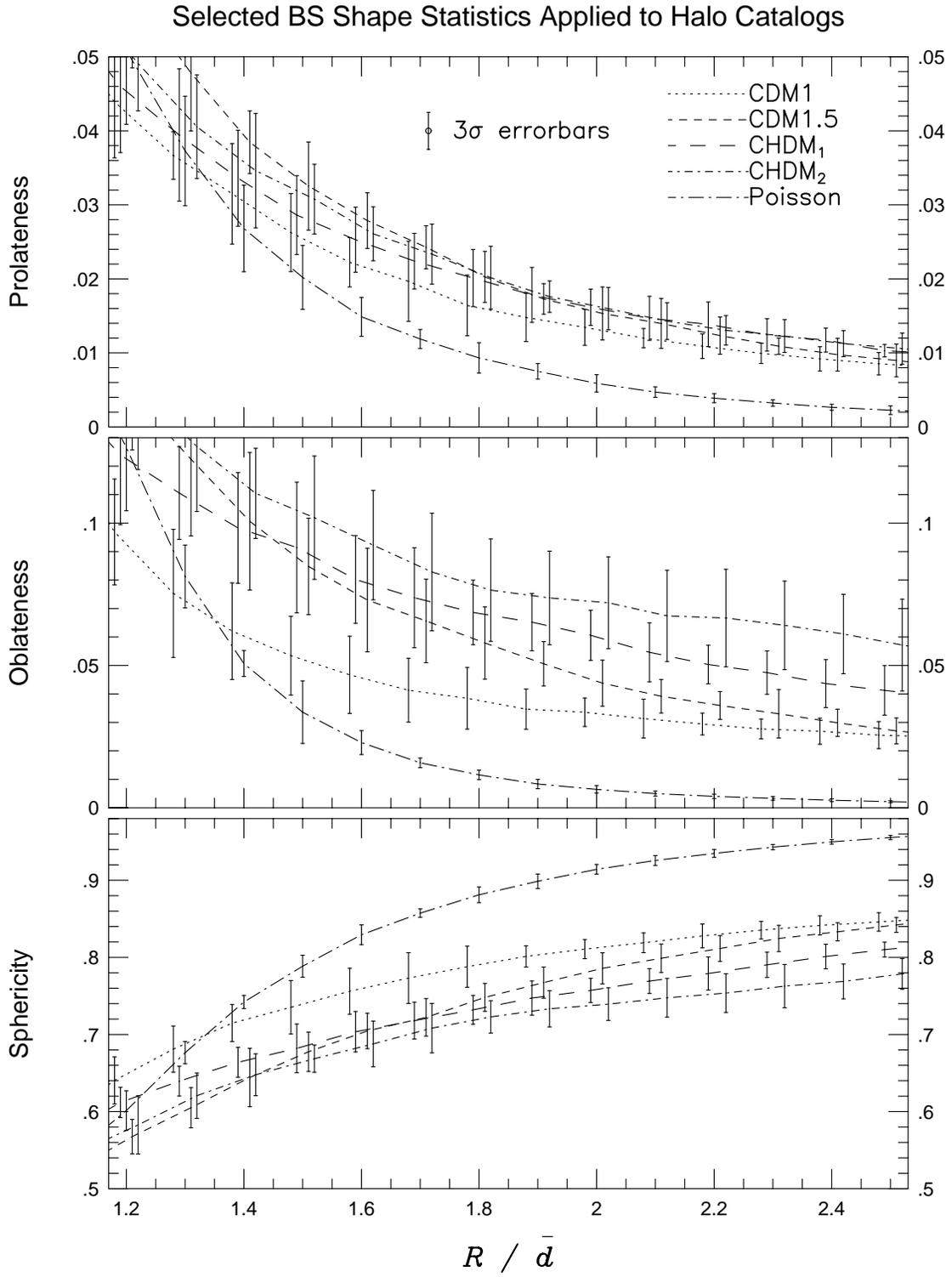}
\caption{Halo Catalogs BS Stats}
\label{fig: gfbubs}
\end{figure}

\bigskip
\noindent Figure~\ref{fig: gfbubs}:
Results from the Babul \& Starkman (1995) statistics 
applied to the halo catalogs after halo breakup.  
Error bars shown are $3\sigma$ resampling errors.
These statistics generally show a very similar behavior
versus each other and versus the Poisson catalog as the $LV$ statistics.
The exception is $BS_{\sl prol}$, which shows very little cosmic variance
for $R/\bar{d} \geq 1.8$.

\begin{figure}
\plotone{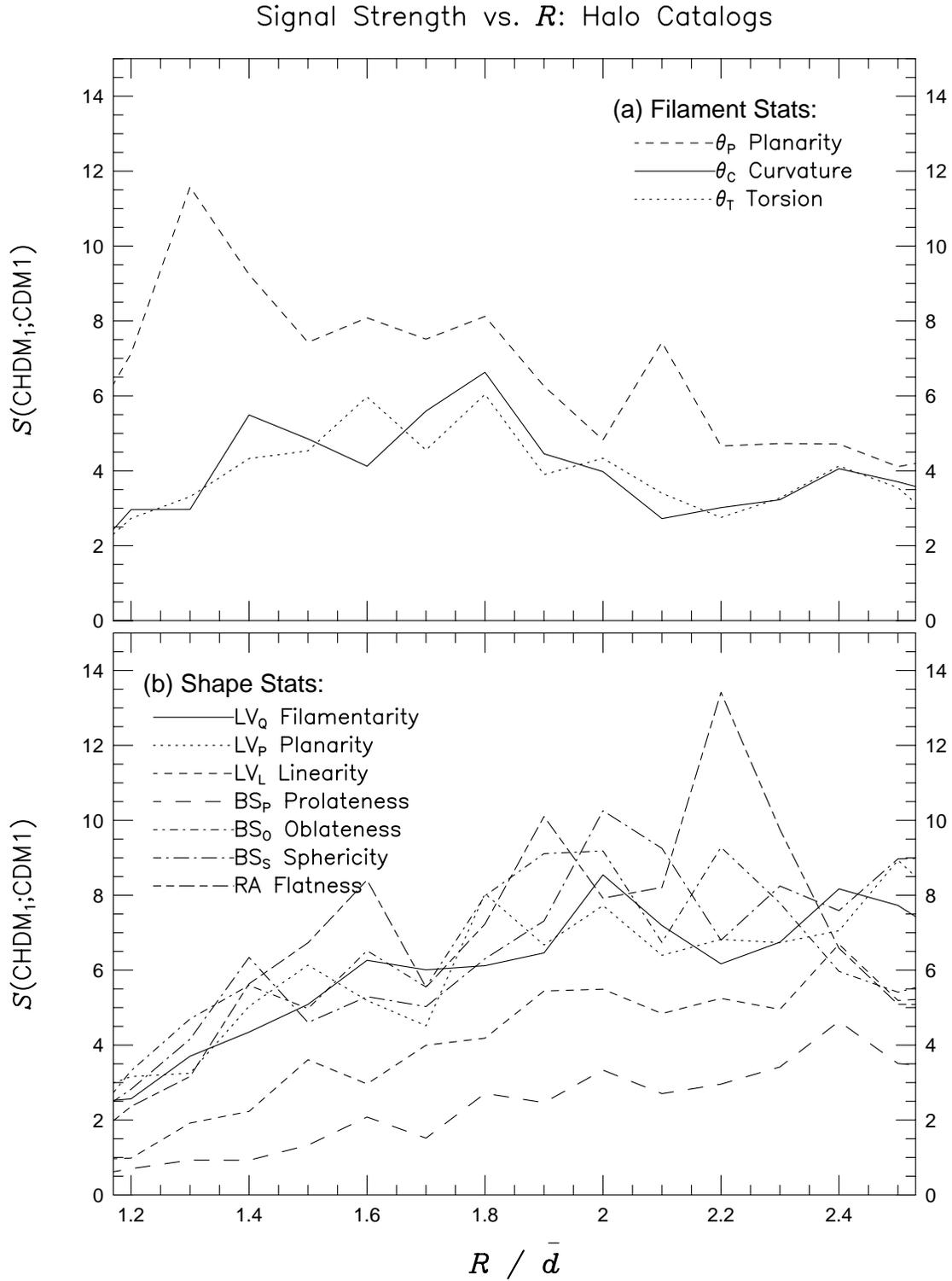}
\caption{Signal strength for statistics applied to halo catalogs after breakup}
\label{fig: dgsig}
\end{figure}

\bigskip
\noindent Figure~\ref{fig: dgsig}.
(a) Signal strengths $S^\theta_{\rm res}$(CHDM$_1$,CDM1), as defined in 
equation~\ref{eq: dgsignal},
for  $\bar\theta_P$, $\bar\theta_C$, and $\bar\theta_T$ applied to 
the halo catalogs.  
All statistics discriminate fairly well, with planarity showing the
most discrimination.
(b) Signal strengths $S^\theta_{\rm res}$(CHDM$_1$,CDM1) for the shape
statistics applied to the halo catalogs.  All statistics show good 
discriminatory power, with the best ones exceeding $\sim 8\sigma$.



\begin{figure}
\plotone{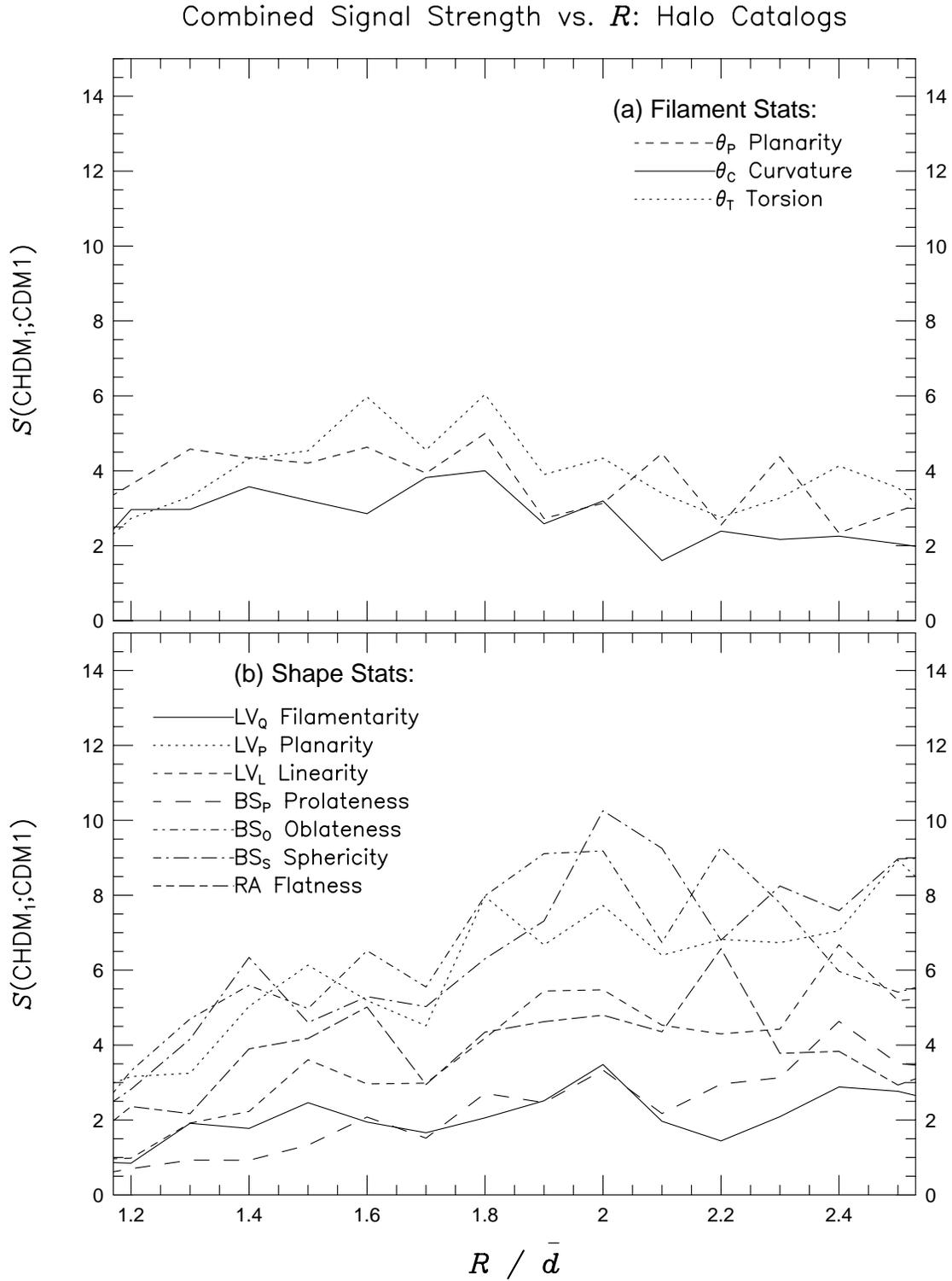}
\caption{Combined signal strength for the halo catalogs}
\label{fig: dgrobful}
\end{figure}

\bigskip
\noindent Figure~\ref{fig: dgrobful}.
Combined signal strengths $S^\theta_{\rm res}$(CHDM$_1$,CDM1)
as defined in equation~\ref{eq: dgrob};
compare to Figure~\ref{fig: dgsig} to see effect of breakup.
(a) $S^\theta_{\rm res+ID}$(CHDM$_1$,CDM1) for the filament statistics
applied to the halo catalogs. 
Comparison with Figure~\ref{fig: dgsig}(a) shows that
breakup causes the most degradation for planarity, some for
curvature, and none for torsion.
(b) $S^\theta_{\rm res+ID}$(CHDM$_1$,CDM1) for the shape statistics
applied to the sky catalogs.  All statistics are quite robust with
respect to halo identification uncertainty, with the exception
of $RA$.

\begin{figure}
\plotone{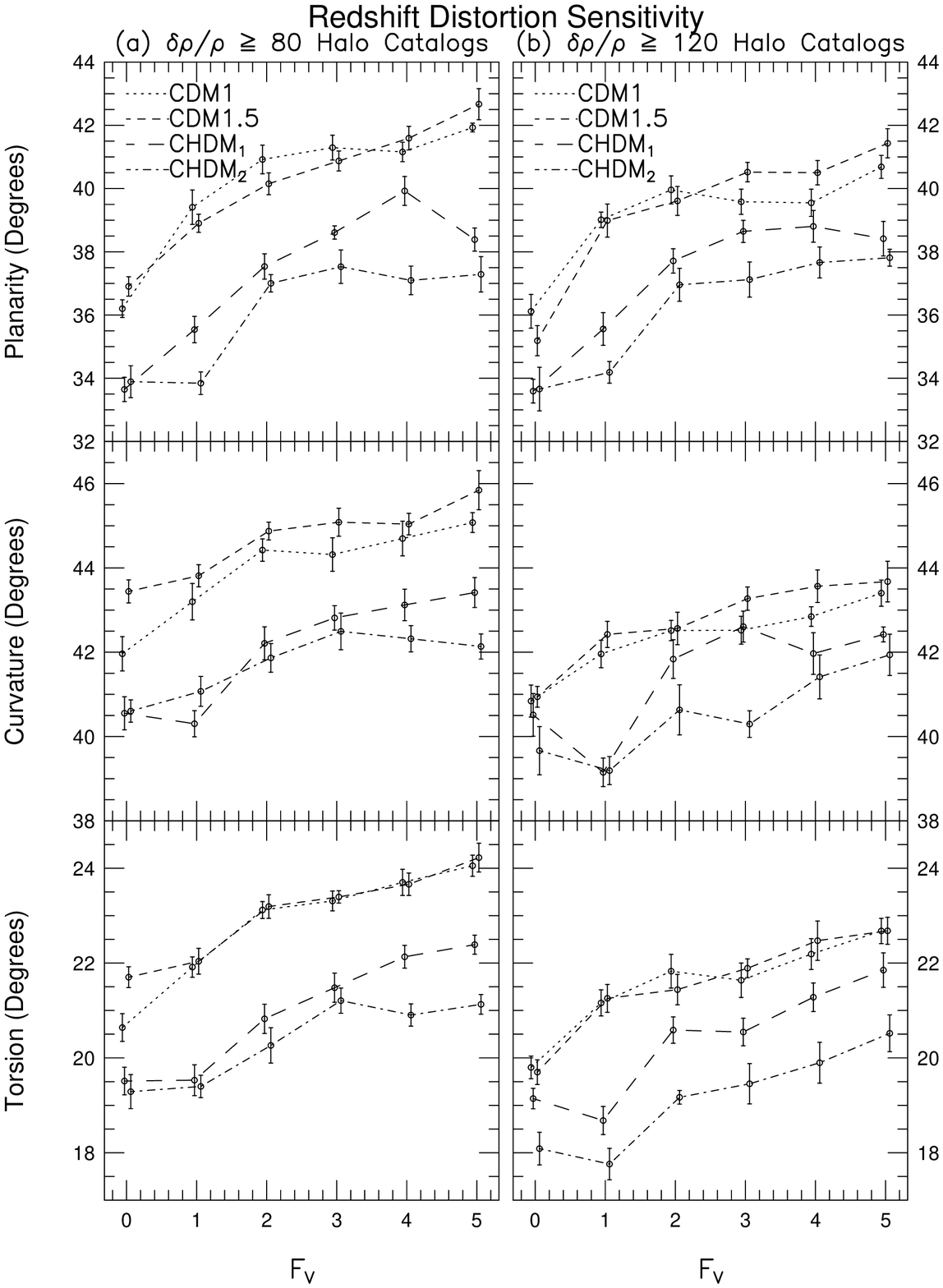}
\caption{Redshift distortion test -- Filament statistics}
\label{fig: reddistfs}
\end{figure}

\bigskip
\noindent Figure~\ref{fig: reddistfs}.
{Filament statistics $\bar\theta_P, \bar\theta_C, \bar\theta_T$ applied
to mock-observed (a) ${\delta\rho/
\rho} > 80$ and (b) $\delta\rho / \rho > 120$ halo catalogs
with velocities scaled from velocity factor $F_V=0$ (real
space) to $F_V=5$ times their actual value, with $R/\bar{d} = 1.5$.
Error bars shown are $1\sigma$ resampling errors.
Going from real
space  ($F_V=0$) to ordinary redshift space ($F_V=1$) decreases
the amount of structure detected, but actually increases discrimination
between models.  
Note: Values for different models are slightly offset in $F_V$ to improve
visibility.}

\begin{figure}
\plotone{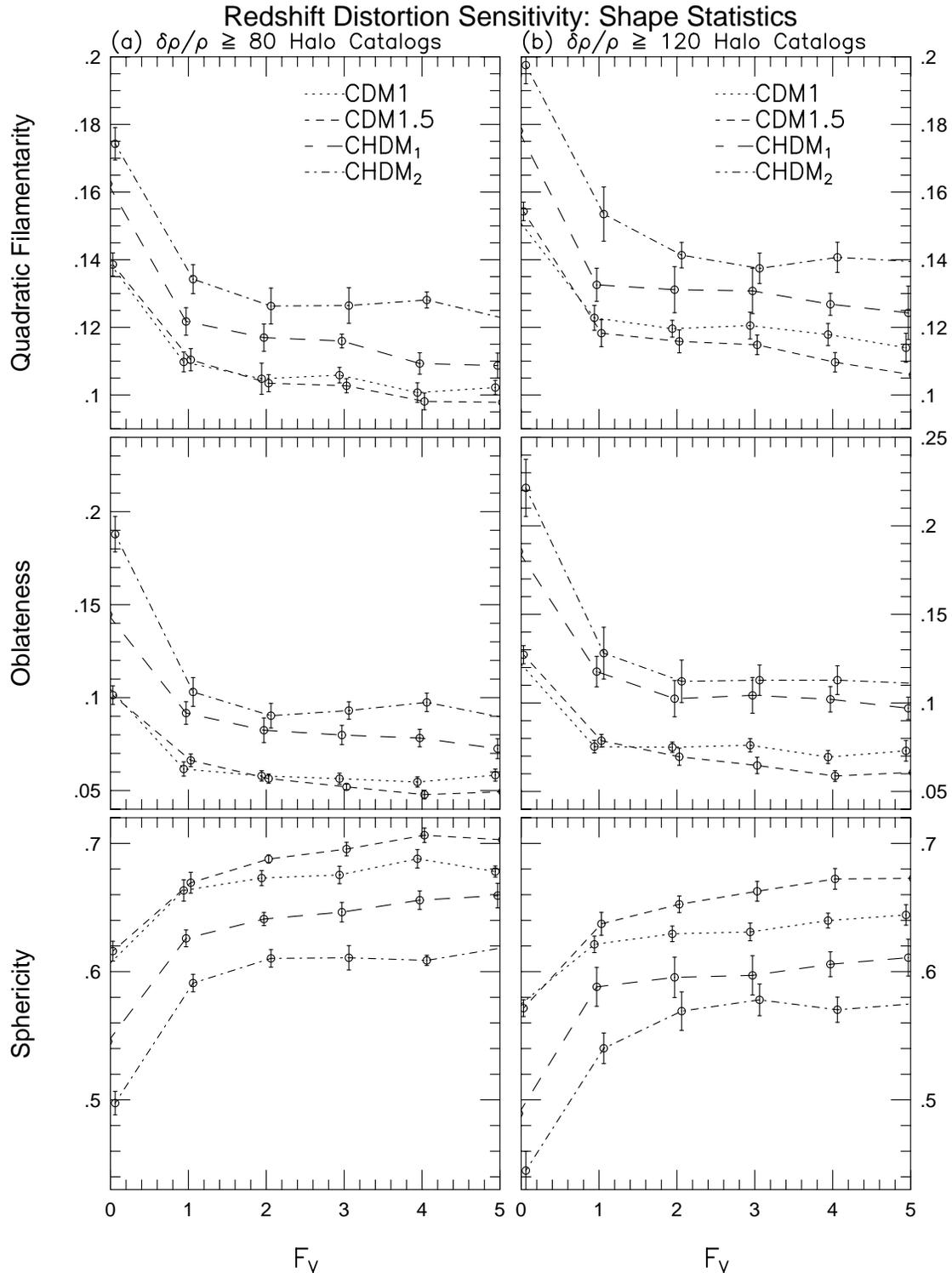}
\caption{Redshift distortion test -- Shape statistics}
\label{fig: reddistshp}
\end{figure}

\bigskip
\noindent Figure~\ref{fig: reddistshp}:
Redshift distortion test applied to catalogs cut at
${\delta\rho\over\rho} \geq 80$ and ${\delta\rho\over\rho} \geq 120$,
then redshifted by their line-of-sight peculiar velocity
multiplied by the velocity scaling factor $F_V$.
Error bars shown are $1\sigma$ resampling erros.
While redshift distortion tends to lower the amount of
struture detected, the discrimination between models is
generally unchanged.

\begin{figure}
\plotone{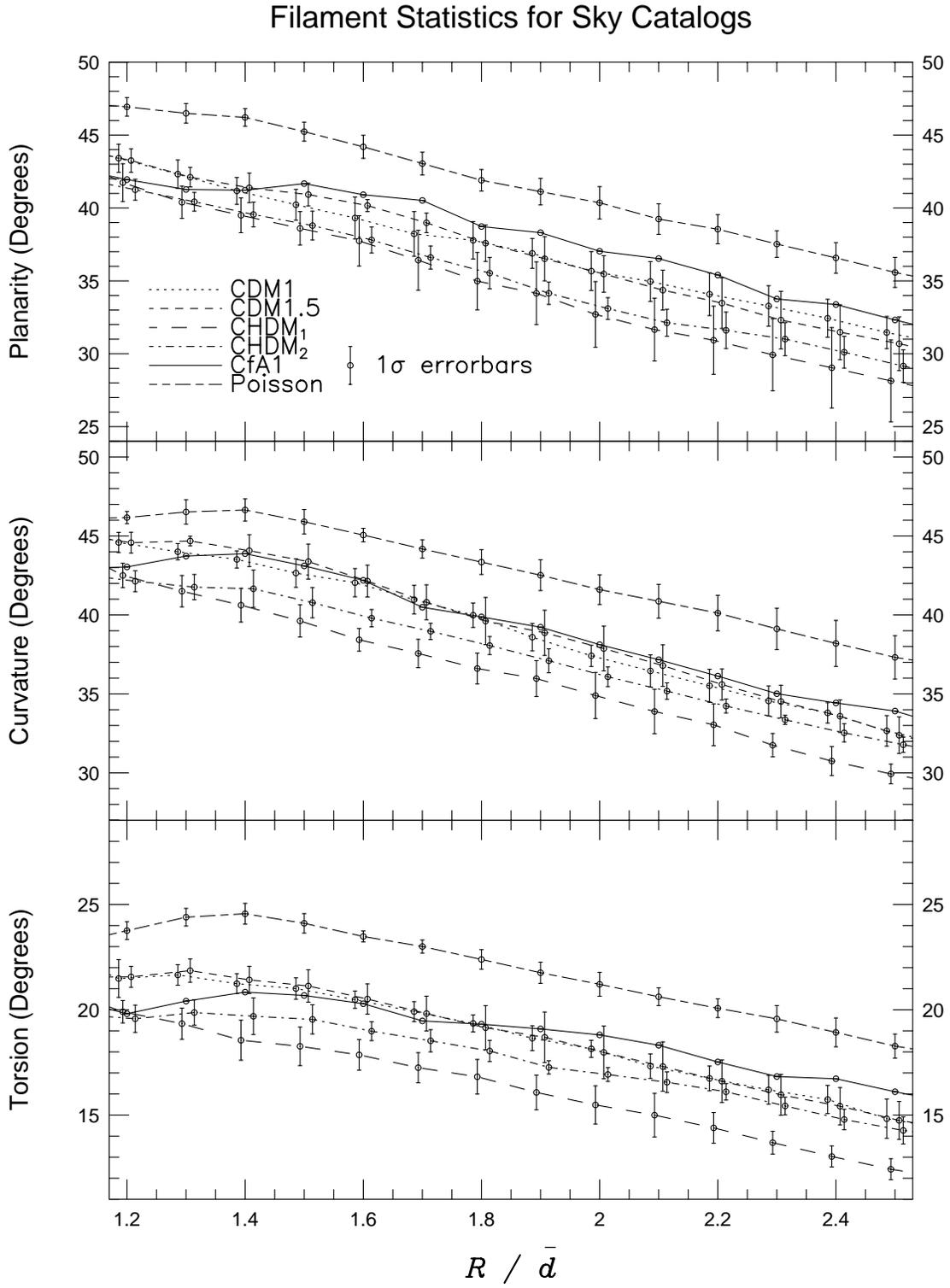}
\caption{Sky Catalog Filament Stats}
\label{fig: dgskystats}
\end{figure}

\bigskip
\noindent Figure~\ref{fig: dgskystats}.
{Filament statistics for the sky catalogs,
versus $R$ in units of ${\bar d (r)}$, the mean interparticle
spacing.  
Error bars shown are $1\sigma$ sky variance errors.
Errors are larger than in the halo catalog statistics
due to sparseness, and Poisson is not as well discriminated from models.  
CDM shows significantly less planarity, curvature, and torsion than CfA1,
while CHDM shows slightly too much.  CfA1 does not
match with any single catalog over all $R$, but does follow
CHDM$_2$ better than the other models, especially for $1.2 \leq R/\bar{d} \leq 2.0$.
The signal-to-noise ratio between CfA1 and Poisson is highest at $R/\bar{d}=1.3$
(note the small Poisson error bar) for all
statistics, indicating optimal sensitivity at this $R$.
Note: Values for different models are slightly offset in $R$ to improve
visibility.}

\begin{figure}
\plotone{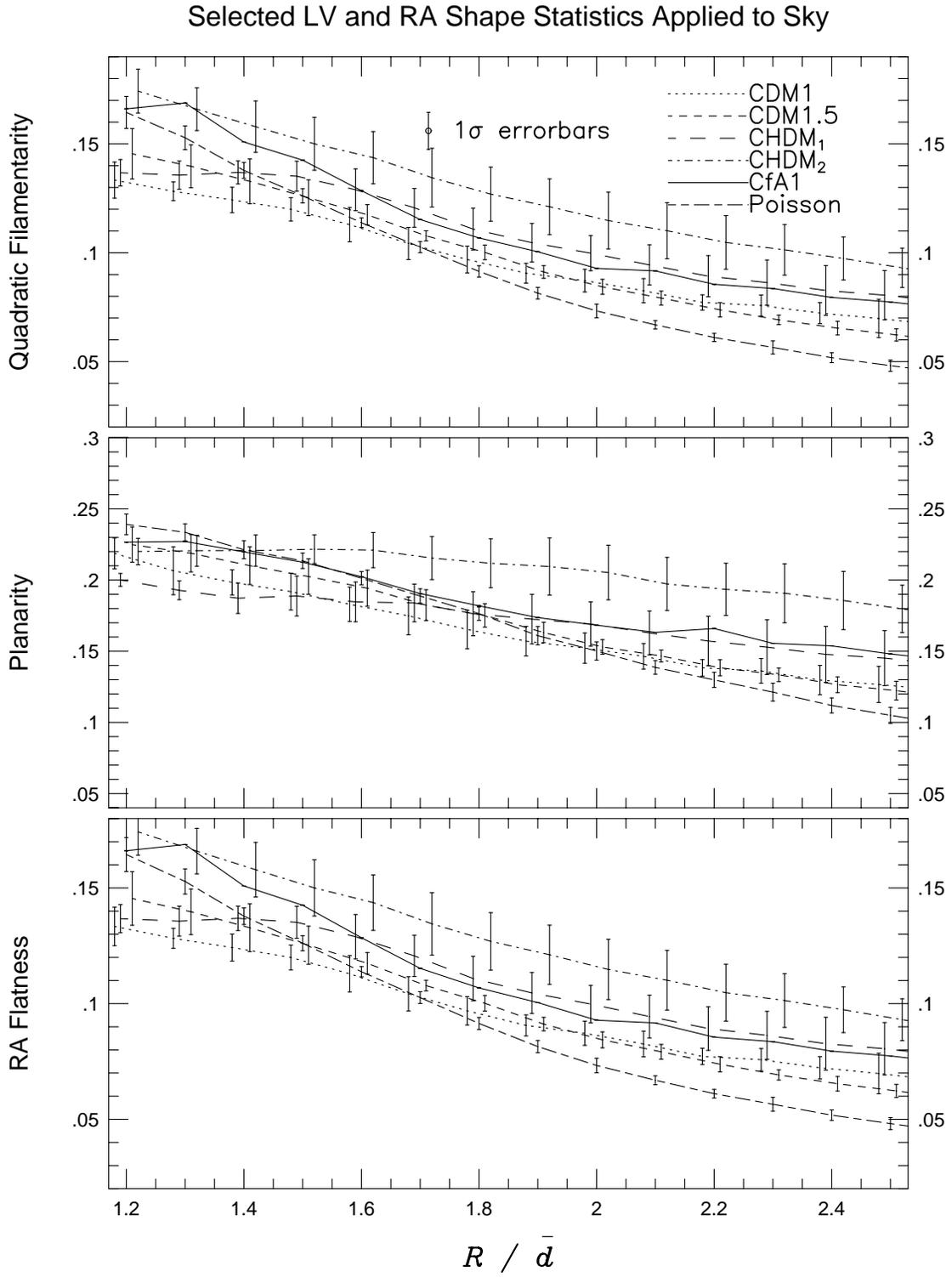}
\caption{Sky Catalog LV \& RA Stats}
\label{fig: gcbulv}
\end{figure}

\bigskip
\noindent Figure~\ref{fig: gcbulv}:
Results from selected Luo \& Vishniac (1995) statistics 
and the Robinson \& Albrecht (1996) statistic applied to the sky catalogs
after halo breakup.  
Error bars shown are $1\sigma$ sky variance errors.
The large errors due to the sparseness of the catalogs
yield a low discrimination between models.  Also, the
Poisson catalog shows significant structure aliasing at all
values of $R/\bar{d}$.  Versus the CfA1 data, no model is ruled out
at more than a $2\sigma$ level, not including cosmic variance.

\begin{figure}
\plotone{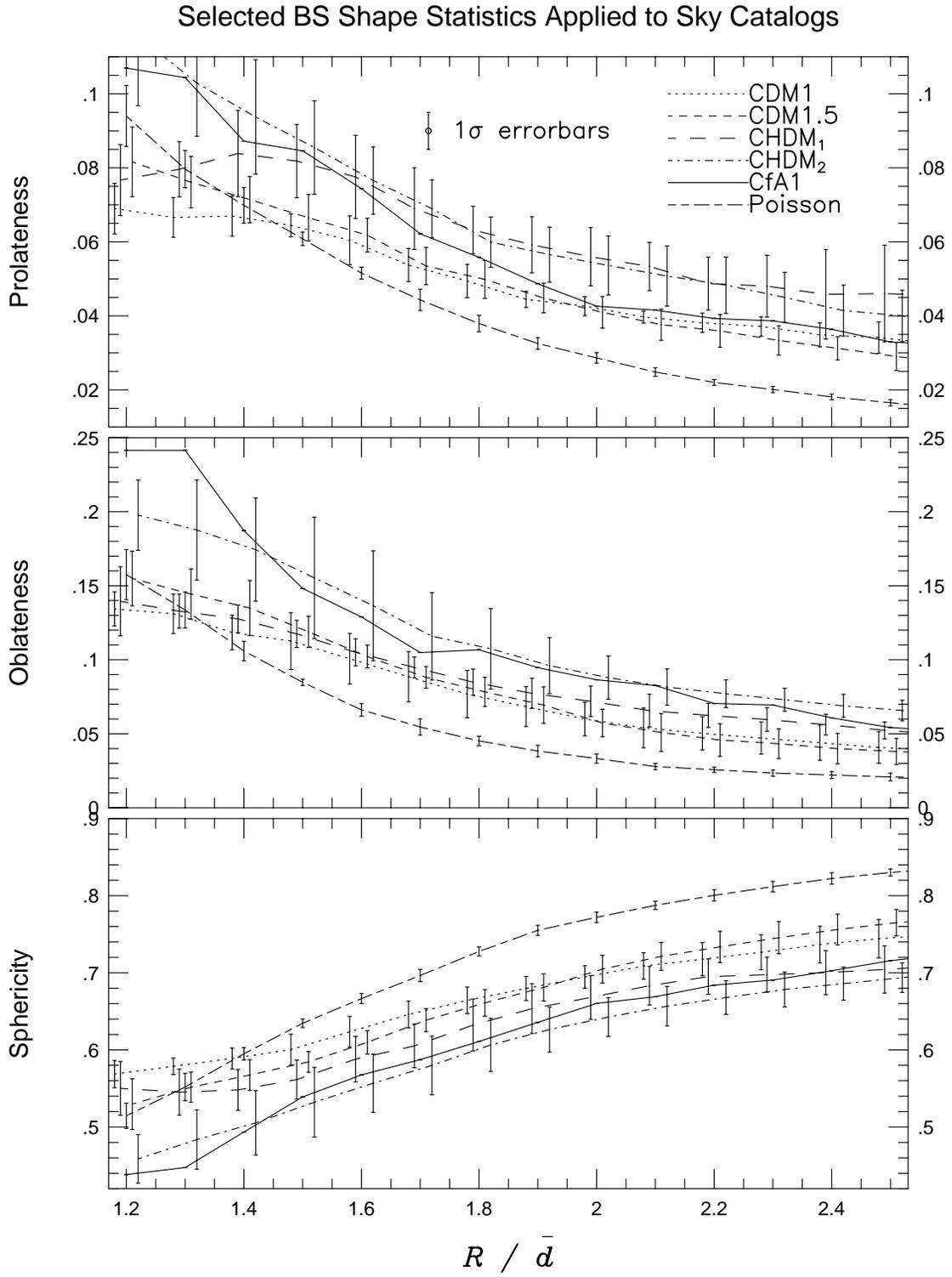}
\caption{Sky Catalog BS Stats}
\label{fig: gcbubs}
\end{figure}

\bigskip
\noindent Figure~\ref{fig: gcbubs}:
Results from the Babul \& Starkman (1995) statistics 
applied to the sky catalogs after halo breakup.
Error bars shown are $1\sigma$ sky variance errors.
The prolateness statistic, which most discriminatory of the
three, also shows low cosmic variance by the crude estimate
of comparing CHDM$_1$ to CHDM$_2$.  The others do not
show good discrimination, and cosmic variance appears to
be larger, although still comparable to sky variance errors.

\begin{figure}
\plotone{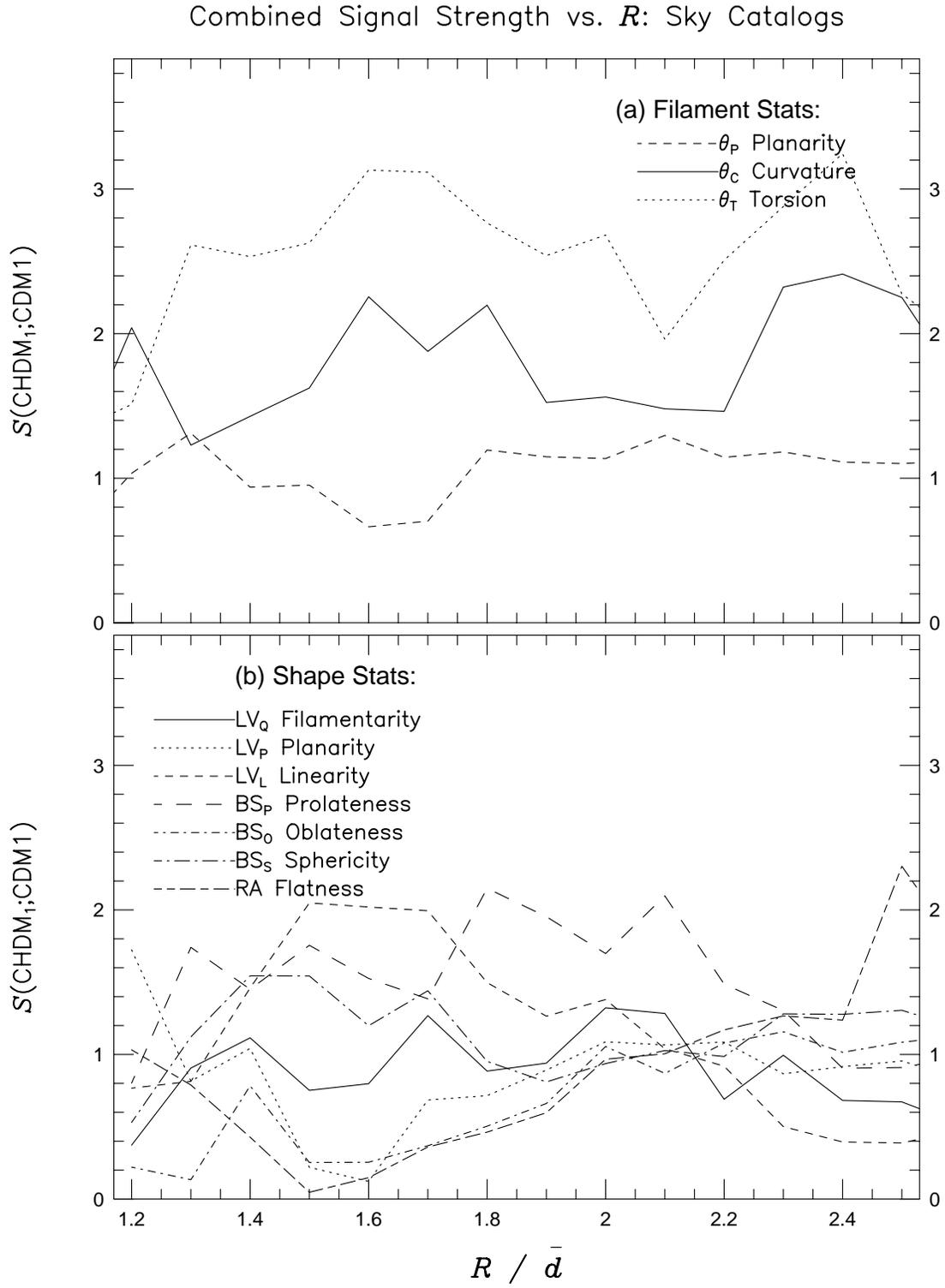}
\caption{Combined signal strength for the sky catalogs}
\label{fig: dgrobcfa}
\end{figure}

\bigskip
\noindent Figure~\ref{fig: dgrobcfa}:
(a) Combined signal strength $S_{\rm sv+id}$(CDM1,CHDM$_1$) for filament 
statistics applied to the CfA1-like sky catalogs.  Torsion clearly shows the
highest discrimination, while curvature also shows some discrimination.
Cosmic variance, specifically excluded in this comparison, is
significant at all $R$.

\noindent
(b) Combined signal strength $S_{\rm sv+id}$(CDM1,CHDM$_1$) for shape 
statistics applied to the sky catalogs.  No statistics shows 
good discriminatory power, with the best one, $BS_{\sl prol}$, 
barely reaching $2\sigma$.


\newpage

\begin{figure}
\plottwo{flowchart.eps}{blank.ps}
\caption{ }
\label{fig: dg_flowchart_filament}
\end{figure}


\end{document}